\RequirePackage[2018-12-01]{latexrelease}

\documentclass{style}
\usepackage{multirow}
\usepackage{adjustbox}
\usepackage{amsmath}
\usepackage{pdflscape}
\usepackage{rotating}
\usepackage{graphicx} 
\usepackage{xcolor}
\usepackage{booktabs}
\usepackage[normalem]{ulem}
\usepackage{hyperref}
\usepackage{cleveref}
\usepackage{amsmath}
\usepackage{ragged2e}
\usepackage[singlelinecheck=false,labelsep=none,justification=justified]{caption}

\DeclareCaptionLabelFormat{customlabel}{Table#1~#2. }
\begin{document}

\title{Decision-informed Neural Networks with Large Language Model Integration for Portfolio Optimization}

\author{Anonymous submission}

\author{{Yoontae Hwang$^{1}$\thanks{Email: yoontae.hwang@eng.ox.ac.uk},
Yaxuan Kong$^{1}$\thanks{Email: yaxuan.kong@eng.ox.ac.uk},
Stefan Zohren$^{1}$\thanks{Corresponding author. Email: stefan.zohren@eng.ox.ac.uk},
Yongjae Lee$^{2}$\thanks{Corresponding author. Email: yongjaelee@unist.ac.kr}}
\affil{$^{1}$University of Oxford \
$^{2}$Ulsan National Institute of Science and Technology(UNIST)}}

\received{Jan 31, 2025}

\maketitle

\begin{abstract}
This paper addresses the critical disconnect between prediction and decision quality in portfolio optimization by integrating Large Language Models (LLMs) with decision-focused learning. We demonstrate both theoretically and empirically that minimizing the prediction error alone leads to suboptimal portfolio decisions. We aim to exploit the representational power of LLMs for investment decisions. An attention mechanism processes asset relationships, temporal dependencies, and macro variables, which are then directly integrated into a portfolio optimization layer. This enables the model to capture complex market dynamics and align predictions with the decision objectives. Extensive experiments on S\&P100 and DOW30 datasets show that our model consistently outperforms state-of-the-art deep learning models. In addition, gradient-based analyses show that our model prioritizes the assets most crucial to decision making, thus mitigating the effects of prediction errors on portfolio performance. These findings underscore the value of integrating decision objectives into predictions for more robust and context-aware portfolio management.

\end{abstract}


\begin{classcode} G11, G17, G45, G61, C53 \end{classcode}

\label{ch:ch1}
\section{Introduction}
The estimation of parameters for portfolio optimization has long been recognized as one of the most challenging aspects of implementing modern portfolio theory \citep{michaud1989markowitz, demiguel2009optimal}. While Markowitz's mean-variance framework \citep{Markowitz1952} provides an elegant theoretical foundation for portfolio selection, its practical implementation has been persistently undermined by estimation errors in input parameters. \citep{chopra1993effect, chung2022effects} demonstrate that small changes in estimated expected returns can lead to dramatic shifts in optimal, while \citep{ledoit2003improved, ledoit2004well} show that traditional sample covariance estimates become unreliable as the number of assets grows relative to the sample size.

These estimation challenges are exacerbated by a fundamental limitation in the conventional approach to portfolio optimization: the reliance on a sequential, two-stage process where parameters are first estimated from historical data, and these estimates are then used as inputs in the optimization problem. This methodology, while computationally convenient, creates a profound disconnect between prediction accuracy and decision quality. Recent evidence suggests that even substantial improvements in predictive accuracy may not translate to better investment decisions. For example, \citep{gu2020empirical,cenesizoglu2012return} demonstrate that while machine learning methods can significantly improve the prediction of asset returns, these improvements do not consistently translate into superior portfolio performance. Similarly, \cite{elmachtoub2022smart} show that this disconnect can lead to substantially suboptimal investment decisions, even when the parameter estimates appear highly accurate by traditional statistical measures. This disconnect raises a profound question: \textit{Are we optimizing for the wrong objective?} The conventional wisdom of training models to minimize prediction error(commonly measured in mean squared error), while ignoring how these predictions influence downstream investment decisions, may be fundamentally flawed.

The significance of this prediction-decision gap has become increasingly acute in modern financial markets, characterized by growing complexity, non-stationary relationships \citep{bekaert2002dynamics}, and regime shifts \citep{guidolin2007asset}. Due to these market dynamics, traditional estimation methods often fail to capture the characteristics of evolving financial relationships. Even modern machine learning approaches struggle to incorporate the complex interplay between macro variables and asset returns \citep{kelly2019characteristics, hwang2024temporal}. The limitations of conventional approaches have created an urgent need for more sophisticated methodologies capable of capturing the intricate dynamics of contemporary financial markets. Recent advances in Large Language Models (LLMs) have introduced promising new directions for addressing these challenges by potentially capturing complex market relationships and incorporating unstructured information. While these approaches have shown remarkable success in time series forecasting across various domains \citep{jin2023time, ansari2024chronos}, their application to financial parameter estimation remains largely unexplored. Moreover, existing attempts to leverage LLMs for financial forecasting continue to follow the traditional sequential approach, focusing solely on prediction accuracy without considering the downstream impact on portfolio decisions \citep{romanko2023chatgpt, nie2024survey}. Despite their enhanced capabilities in capturing patterns in data, LLMs alone do not address the core problem of bridging the gap between predictive modeling and optimal portfolio construction.

A more comprehensive approach is needed to integrate prediction and optimization stages in portfolio management. A more promising direction lies in decision-focused learning frameworks \citep{mandi2024decision}, which represent a significant departure from traditional approaches by directly integrating the prediction and optimization stages. The emergence of decision-focused learning frameworks represents a significant departure from this approach by integrating the prediction and optimization stages. While these frameworks have shown promise in combinatorial optimization problems \citep{amos2017optnet, cvxpylayers2019}, their application to portfolio optimization has been limited. Early attempts in finance have primarily focused on simple linear models \citep{butler2023integrating}, leaving the potential of decision-focused learning in complex portfolio optimization largely untapped. For instance, \cite{elmachtoub2022smart} demonstrate the benefits of integration in linear optimization problems, but extending these insights to the non-linear, dynamic nature of portfolio optimization remains a significant challenge. The development of techniques for differentiating through convex optimization problems \citep{amos2017optnet, cvxpylayers2019} has created theoretical possibilities for more sophisticated applications, yet their practical implementation in portfolio management continues to face substantial computational and methodological challenges.

This paper proposes a framework that bridges the gap between advanced representation learning and decision-focused optimization in portfolio management. Our approach integrates the representational power of LLMs with the principles of decision-focused learning, creating an decision informed neural network architecture that simultaneously captures complex market relationships and optimizes for portfolio decisions. By developing attention mechanisms that incorporate both cross-sectional asset relationships and temporal dependencies, we enable the model to learn representations that are both predictively accurate and decision-aware. Furthermore, our proposed loss function simultaneously optimizes for statistical accuracy and portfolio performance, ensuring the model's predictions directly translate into better investment decisions.

The main contributions of this paper are as follows:
\begin{itemize}
\item We develop a decision-informed neural network framework that integrates LLMs for portfolio optimization. Our approach carefully considers the unique characteristics of financial markets by incorporating multiple data dimensions: cross-sectional relationships between assets, temporal market dynamics, and the influence of macroeconomic variables. This comprehensive modeling approach ensures that the LLM's powerful representation capabilities are properly adapted to the specific challenges of portfolio optimization.
\item We introduce an attention mechanism that selectively processes three crucial aspects of financial markets through learned representations from Large Language Models (LLMs): asset-to-asset relationships, temporal dependencies, and external macro variables. This mechanism implements an efficient filtering strategy that identifies and extracts only the most relevant information from the rich LLM representations, significantly reducing computational overhead while preserving essential market insights.  By selectively attending to the most relevant features within each aspect, our model achieves both superior computational efficiency and enhanced interpretability in capturing complex market interactions.
\item We propose a hybrid loss function that bridges the gap between statistical prediction accuracy and decision-focused learning for portfolio optimization objectives. This function combines traditional prediction metrics with portfolio performance measures, ensuring that the model learns parameters that are both statistically sound and economically meaningful. Our approach directly addresses the prediction-decision gap while maintaining the model's ability to capture complex market relationships. To the best of our knowledge, this is the first study to combine LLM and DFL from a portfolio optimization perspective.

\end{itemize}


\label{ch:ch2}
\section{Related work}
In this section, we review two key research streams relevant to our work: parameter estimation in portfolio optimization and deep learning applications in financial forecasting. The first stream examines traditional and robust estimation techniques, while the second explores how modern machine learning approaches have transformed financial prediction. 

\subsection{Parameter estimation in portfolio optimization}
The foundation of modern portfolio theory rests upon accurate parameter estimation, particularly for expected returns and covariance matrices. Since \cite{Markowitz1952} seminal work establishing the mean-variance optimization framework, researchers have grappled with the challenge of reliably estimating these crucial parameters from historical data \cite{tan2020estimation, firoozye2023canonical}. This challenge has become central to the field of quantitative finance, as the performance of optimal portfolios heavily depends on the quality of these estimates.

The extensive reliance on historical financial data for parameter estimation has proven instrumental in advancing modern financial theory and practice. Historical data provides the empirical foundation for estimating critical parameters including expected returns, volatility, and covariance matrices—essential inputs that drive portfolio optimization, risk management, and asset pricing models. This approach has led to breakthrough developments in financial modeling, most notably the Capital Asset Pricing Model (CAPM) \citep{sharpe1964capital, lintner1975valuation} and the Fama-French factor models \citep{fama1993common,fama2015five}. This mean-variance optimization framework, however, revealed significant challenges in parameter estimation. The sensitivity of portfolio optimization to parameter estimates was first systematically documented by \cite{michaud1989markowitz}, who characterized mean-variance optimization as "error maximization." This insight was further developed by \cite{best1991sensitivity}, who demonstrated the hypersensitivity of optimal portfolio weights to changes in mean estimates. \citep{chopra1993effect, chung2022effects} provided crucial quantitative evidence, showing that errors in mean estimates have approximately ten times the impact of errors in variance estimates on portfolio performance.

In response to these challenges, researchers developed increasingly sophisticated estimation techniques. Early efforts focused primarily on improving covariance matrix estimation. \citep{chan1999portfolio, loffler2003effects} proposed utilizing high-frequency data for enhanced volatility forecasts, while \citep{jagannathan2003risk} made the crucial observation that imposing portfolio constraints could effectively shrink extreme covariance estimates. The recognition of parameter uncertainty led to more sophisticated approaches, such as the shrinkage method \citep{ledoit2003improved, ledoit2004well,kourtis2012parameter}, which combines sample estimates with structured estimators to reduce estimation error. As understanding of estimation challenges deepened, robust estimators gained prominence, including the minimum covariance determinant (MCD) estimator \citep{rousseeuw1999fast} and the minimum volume ellipsoid (MVE) estimator \citep{van2009minimum}. A significant advancement came with \cite{gerber2022gerber} introduction of the Gerber statistic, a robust co-movement measure that extends Kendall's Tau by capturing meaningful co-movements while remaining insensitive to extreme values and noise.

The emergence of machine learning has transformed the landscape of parameter estimation \citep{kim2021mean, kim2024overview}. While traditional machine learning approaches typically treated prediction and optimization as separate steps, recent research has explored more integrated approaches. Notable contributions include the works of \cite{ban2018machine} and \cite{feng2020taming}, who demonstrated significant improvements over traditional approaches. However, these studies relied on the Sharpe ratio, which is an indirect performance measure calculated from returns and volatility after portfolio construction, rather than directly obtaining optimal portfolio weights through the optimization process itself. This indirect approach may not fully capture the actual decision-making process inherent in portfolio optimization, where the primary goal is to determine optimal portfolio weights that satisfy specific investment objectives and constraints. Fortunately, technological advances particularly in differentiable optimization have opened new frontiers. The introduction of cvxpylayers \citep{cvxpylayers2019} and the work of \cite{amos2017optnet} on differentiable optimization layers have enabled end-to-end training of machine learning models that incorporate the portfolio optimization step directly into the parameter estimation process. While these approaches represent significant progress in bridging the gap between prediction and optimization, current implementations often rely on simplistic linear models that may not fully capture the complex, non-linear dynamics of financial markets \citep{costa2023distributionally, anis2025end}. Moreover, these models typically focus on a limited set of financial variables, potentially overlooking important external factors such as macroeconomic condition, stock and sector relationship that can significantly impact portfolio performance. See \cite{lee2024overview} for more detailed review of the evolution from traditional two-stage approaches to modern end-to-end learning frameworks, including decision-focused learning (DFL) methodologies.

\subsection{Time-series forecasting with deep learning}

The application of deep learning to financial time-series forecasting represents a significant advancement in addressing the parameter estimation challenges. While traditional approaches to parameter estimation often struggle with the complex, non-linear relationships inherent in financial markets, deep learning models have demonstrated remarkable capability in capturing these dynamics. However, as discussed in our examination of parameter estimation challenges, improved predictive accuracy does not necessarily translate to better portfolio decisions.

Recent advances in deep learning architectures, particularly those based on attention mechanisms, have revolutionized time-series forecasting across various domains. The Transformer architecture and its variants have achieved state-of-the-art performance in various domains through innovations in handling long sequences \citep{informer_2021}, capturing interactions between different time scales \citep{zhou2022fedformer}, and modeling temporal patterns \citep{wu2023timesnet}. Unlike traditional autoregressive predictors such as LSTM, transformer-based models employ \textit{generative-style decoders} as non-autoregressive predictors, facilitating more efficient time series prediction. Notable advances include the Crossformer architecture \citep{zhang2023crossformer}, which explicitly models cross-dimensional dependencies, and PatchTST \citep{Yuqietal_PatchTST}, which adapts vision Transformer techniques to time-series data. The iTransformer \citep{liu2023itransformer} further enhances this approach by treating the temporal dimension as channels, enabling more efficient processing of long sequences. Also, The emergence of Large Language Models (LLMs) has introduced new possibilities for forecasting. Recent works such as Chronos \citep{ansari2024chronos} and GPT4TS \citep{zhou2023one} demonstrate that LLMs can effectively capture complex temporal patterns while incorporating broader market context. The PAttn framework \citep{tan2024language} specifically addresses various forecasting problems by combining linguistic and numerical features in a unified architecture.

However, these advances in predictive modeling, while impressive, often fall short in addressing the fundamental challenges of portfolio optimization. The primary limitation lies in their focus on minimizing prediction error rather than optimizing investment decisions. Even when these models incorporate financial performance metrics like the Sharpe ratio into their loss functions, they typically do so in a manner that fails to capture the full complexity of the portfolio optimization problem. This disconnect becomes particularly apparent when considering the challenges identified in our parameter estimation analysis \citep{hwang2024temporal}. While deep learning models may achieve superior accuracy in forecasting individual asset returns or volatilities, they often fail to account for the complex interplay between estimation errors and portfolio weights that makes the parameter estimation problem so challenging. The sensitivity of optimal portfolio weights to small changes in input parameters, as demonstrated by \citep{chopra1993effect}, suggests that even highly accurate predictions may lead to suboptimal portfolio decisions if the prediction-optimization interface is not properly considered. 

The application of general-purpose time-series models to financial markets presents additional challenges beyond prediction accuracy. While models like TimesNet \citep{wu2023timesnet} and Fedformer \citep{zhou2022fedformer} excel in capturing temporal dependencies, they often struggle to incorporate broader macroeconomic factors and market conditions that significantly influence asset prices. These models typically focus on historical price patterns while failing to account for important external factors such as monetary policy changes, real estate prices, or shifts in market sentiment. This limitation extends beyond individual models, as most existing research has focused primarily on pattern recognition within historical data. A more promising direction may be integrating deep learning models with methods that can effectively incorporate broader market context and macroeconomic indicators.

The evolution of deep learning approaches in financial time-series forecasting thus mirrors the broader challenges in portfolio optimization: while technical capabilities continue to advance, the fundamental challenge lies not in improving predictive accuracy, but in developing frameworks that directly optimize for investment decisions. This observation reinforces our motivation for developing more integrated approaches that combine the representational power of modern deep learning architectures with explicit consideration of the portfolio optimization objective.


\label{ch:ch3}
\begin{figure}[h!] 
  \centering
  \includegraphics[width=\linewidth]{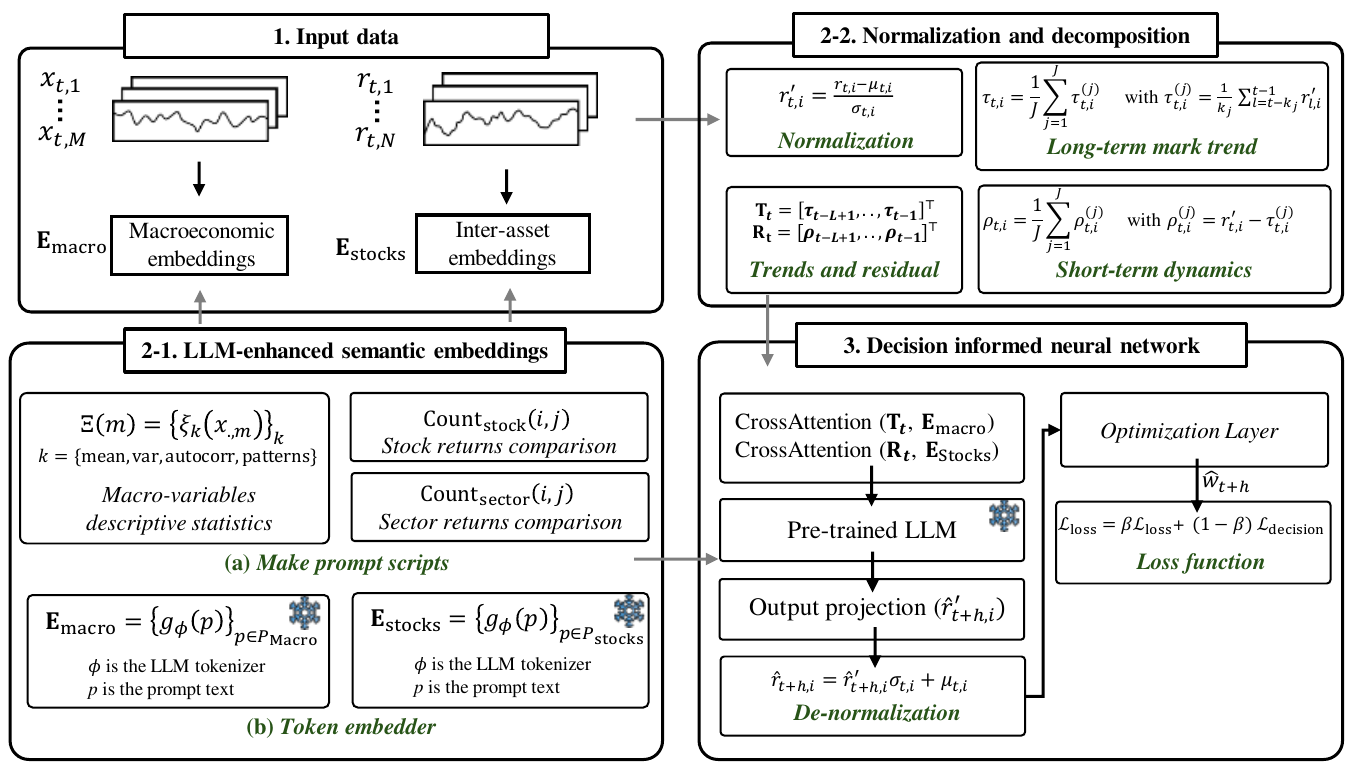}
   \captionsetup{font=footnotesize}
   \caption{Schematic of the proposed Decision-Informed Neural Network (DINN) architecture for unified return forecasting and portfolio selection. The entire system is trained end-to-end to align predictive accuracy with decision quality.}
  \label{fig:model}
\end{figure}

\section{Decision-informed neural networks with large language model integration for portfolio optimization (DINN)}
We introduce a Decision-Informed Neural Network (DINN) that unifies forecasting and portfolio selection within a single learning framework. Unlike traditional methods that treat return prediction and portfolio optimization as separate tasks, DINN merges them via three key components. First, an \emph{input embedding process} captures market dynamics and semantic relationships using LLM-based representations, ensuring that both numeric time series and textual context inform the model. Second, a \emph{cross-attention mechanism} fuses these diverse inputs into coherent return forecasts, allowing interactions between multiple data modalities. Finally, a \emph{differentiable optimization layer} uses these forecasts to produce optimal portfolio weights, enabling the model to refine both predictive accuracy and decision quality simultaneously.  An overview of the DINN architecture is illustrated in \Cref{fig:model}. By jointly training all components, DINN directly aligns return predictions with end-to-end portfolio performance.

\subsection{Preliminaries}
Throughout this paper, let $N \in \mathbb{N}$ denote the number of risky assets in the portfolio. We consider a discrete-time financial market with a finite horizon $ T \in \mathbb{N} $. Let $\{r_{t}\}_{t=1}^{T}$ be a sequence of asset excess returns, where $r_{t} \in \mathbb{R}^{N}$ and $r_{t,i}$ is the excess return of asset $i$ at time $t$. Let $r_{t,i} = \frac{P_{t,i} - P_{t-1,i}}{P_{t-1,i}} - r_{f}$, where $P_{t,i}$ denotes the price of asset $i$ at time $t$ and $r_{f}$ is the risk-free rate. We assume $r_f$ is a known, time-invariant constant. Define $w_{t} \in \mathbb{R}^{N}$ as the portfolio weight vector at time $t$. Let $\mathcal{W} \subseteq \mathbb{R}^{N}$ be the feasible set of portfolio weights, for example $\mathcal{W} = \{ w \in \mathbb{R}^{N}: w_i \geq 0,\ \sum_{i=1}^{N} w_i = 1 \}$. For a given lookback length $ L \in \mathbb{N} $, consider historical returns and macroeconomic variables over the period $\{t-L, \ldots, t-1\}$: 

\begin{equation} r_{t-L: t-1}=\left(r_{t-L}, \ldots, r_{t-1}\right) \in \mathbb{R}^{L \times N},
            \quad x_{t-L: t-1}=\left(x_{t-L}, \ldots, x_{t-1}\right) \in \mathbb{R}^{L \times M}
\end{equation} where $x_{t} \in \mathbb{R}^{M}$ encapsulates $M$ macroeconomic features observed at time $t$.

We consider a forecasting model $ f_{\theta} $ parameterized by $\theta$. This model, given historical data and macroeconomic features, produces predicted returns $\hat{r}_{t+1:t+H} = (\hat{r}_{t+1}, \hat{r}_{t+2}, \ldots, \hat{r}_{t+H})$ but also of the corresponding predicted portfolio weights ${\hat{w}}_{t+1:t+H} = ({\hat{w}}_{t+1}, {\hat{w}}_{t+2}, \ldots, {w}_{t+H})$ over a forecast horizon $ H \in \mathbb{N} $. Formally, we have: 
\begin{equation} 
{\hat{r}}_{t+1: t+H}=f_{\theta}\left(r_{t-L: t-1}, x_{t-L: t-1}\right) 
\end{equation}
where $\widehat{r}_{t+1:t+H} = \bigl(\widehat{r}_{t+1},\,\ldots,\,\widehat{r}_{t+H}\bigr)\in \mathbb{R}^{H\times N}$. The vector $\widehat{r}_{t+h}\in\mathbb{R}^{N}$ thus represents the \emph{predicted} excess returns for the $N$ assets at time $t+h$.

Once $\hat{r}_{t+1 : t+H}$ is obtained, the corresponding portfolio weights $\hat{w}_{t+1 : t+H} 
= \bigl(\hat{w}_{t+1}, \ldots, \hat{w}_{t+H}\bigr)$ are determined by solving a suitable optimization problem that incorporates risk-return trade-offs over the forecast horizon $H$. We defer the precise formulation to Section~3.3.3. The predicted returns $\hat{r}_{t+h}$ act as inputs to a differentiable optimization layer that selects an \emph{optimal} allocation $\hat{w}_{t+h}$ to balance risk and reward under model predictions.

If we hypothetically had complete knowledge of the future, i.e.\ the ``true'' future returns $r_{t+H}^{\star}$, true expected returns $\mu_{t+H}^{\star}$, and covariance $\Sigma_{t+H}^{\star}$, we could compute the \emph{ex-post} optimal portfolio weights $w_{t+H}^{\star}$ by substituting the actual (rather than predicted) parameters into the same portfolio optimization problem. Formally:
\begin{equation}
    w_{t+H}^{\star} 
    = \arg\min_{w \in \mathcal{W}}
    \Bigl[\lambda\,\bigl\|L_{t+H}^{\star}\,w \bigr\|_2 
    \;-\; (\mu_{t+H}^{\star})^{\top} w 
    \Bigr] 
    \quad \text{where} \quad
    \Sigma_{t+H}^{\star} 
    = L_{t+H}^{\star}\,\bigl(L_{t+H}^{\star}\bigr)^{\top}.
\end{equation}

The difference between the performance of $\hat{w}_{t+H}$ (obtained via predicted returns) and $w_{t+H}^{\star}$ (with full foresight) will later be used to evaluate the ``decision quality'' of the forecasting pipeline. Although we distinguish between predicted returns $\hat{r}_{t+1:t+H}$ and the corresponding weights $\hat{w}_{t+1:t+H}$ for notational clarity, the DINN framework integrates these components into a unified, end-to-end pipeline. That is, the forecast of $\hat{r}_{t+1:t+H}$ directly informs the subsequent portfolio choice, and the model is trained with awareness that its predictions will drive the ultimate decision.

\subsection{Input embeddings}
Our input embedding process is designed to systematically incorporate temporal patterns, asset interactions, and textual context before generating forecasts. First, we \emph{normalize time series data} to stabilize training and ensure comparability across assets. Next, \emph{kernel-based trend-residual decompositions} separate persistent market trends from shorter-term fluctuations, highlighting both low-frequency and high-frequency signals. Finally, \emph{LLM-enhanced semantic embeddings} integrate sector-level yields and pairwise asset relationships into the model, thereby capturing broader economic and inter-asset context. These structured embeddings may provide a strong foundation for subsequent attention-based modeling and decision-focused optimization.

\subsubsection{Time-series normalization and decomposition} We begin by transforming the raw input data into a structured representation well-suited for accurate forecasting and decision-focused optimization. Let $\{r_{t}\}_{t=1}^{T}$ be a sequence of excess returns for $N$ assets, where $r_{t} \in \mathbb{R}^{N}$. To ensure numerical stability and promote effective learning, we first normalize the historical returns. For each asset $i \in \{1, \ldots, N \}$ over a lookback window of length $L$, define the sample mean $\mu_i$, and standard deviation $\sigma_i$ as $\mu_{t, i} = \frac{1}{L}\sum_{k=t-L}^{t-1}r_{k,i}$, $\sigma_{t, i} = \sqrt{\frac{1}{L}\sum_{k=t-L}^{t-1}(r_{k,i}-\mu_{t,i})^{2}+\epsilon}$, respectively. Here $\epsilon>0$ is a small constant to avoid division by zero. And then, we can get the normalized returns (i.e., $r_{t,i}^{\prime} = \frac{r_{t,i}-\mu_{t,i}}{\sigma_{t,i}}$). This normalization step\citep{kim2021reversible} ensures that differences among assets are measured relative to their historical scales, improving training stability and preventing certain assets from dominating the optimization process solely due to larger raw magnitudes.

Next, we apply a multi-scale decomposition \citep{wu2021autoformer, zhou2022fedformer} to the normalized returns to capture both persistent trends and transient fluctuations. Let $\{k_{j}\}_{j=1}^{J}$ be a collection of kernel sizes. For each $j$, we can define as $\tau_{t,i}^{(j)} = \frac{1}{k_j}\sum_{\ell=t-k_j}^{t-1}r_{\ell, i}^{\prime}$, $\rho_{t,i }^{(j)} = r_{t,i}^{\prime}-\tau_{t,i}^{(j)}$. By aggregating across all scales, we can obtain 
\begin{equation}
    \tau_{t,i} := \frac{1}{J}\sum_{j=1}^{J}\tau_{t,i}^{(j)}, \quad \rho_{t,i} := \frac{1}{J}\sum_{j=1}^{J}\rho_{t,i}^{(j)}
\end{equation}
This approach allows the model to focus separately on the long-term market trend (captured by $\tau_t$) and short-term dynamics (captured by $\rho_t$), where $\rho_t$ represents the remaining variations after extracting the trend component, potentially enhancing forecasting accuracy and stability.

\subsubsection{LLM-enhanced semantic embeddings} While normalized and decomposed returns offer valuable insights into market structures, their representational capacity can be significantly enriched by incorporating Large Language Model (LLM)-based embeddings \citep{zhou2023onefitsall, jin2023time, cao2024tempo}. To achieve this, we integrate two distinct types of LLM-based embeddings: one capturing inter-asset relationships, and another encoding macroeconomic information.

\textbf{Inter-asset embeddings:}  Consider a set of assets indexed by $ i \in \{1,\ldots,N\} $, each mapped to a sector $ S(i) $ drawn from a finite set $\mathcal{S}$. Using large language model (LLM)-based textual descriptions, we establish a mapping from each asset to its corresponding sector. Once this mapping is determined, we construct sector-level returns over a historical lookback period to complement asset-level historical returns.

More specifically, let $ L \in \mathbb{N} $ be the lookback length, and consider the historical returns $\{r_{t,i}\}_{t=t-L}^{t-1}$ for each asset $ i $. The sector-level yield at time $ u \in \{t-L,\ldots,t-1\} $ for a sector $ s \in \mathcal{S} $ is defined as:
\begin{equation}
    r_{u,s}^{\text{sector}} = \frac{1}{|\mathcal{A}(s)|} \sum_{i \in \mathcal{A}(s)} r_{u,i}    
\end{equation}

where $\mathcal{A}(s) = \{ i \in \{1,\ldots,N\} : S(i)=s \}$. This produces, for each sector, a time series $\{r_{u,s}^{\text{sector}}\}_{u=t-L}^{t-1}$ that may reveal common patterns, systemic shifts, or sectoral performance trends during the lookback window. 

Next, to capture direct relationships among individual assets, for each pair $(i,j)$ with $i \neq j$, we measure relative historical performance by counting how frequently one asset outperforms the other:  
\begin{equation}
    \text{Count}_{\text{stock}}(i,j) 
    = \Bigl|\bigl\{\, u \in [t-L,t-1] \colon r_{u,i} > r_{u,j}\bigr\}\Bigr|.
    \end{equation}

Similarly, we define a sector-level outperformance count to capture how often the sector of asset $i$ outperforms the sector of asset $j$:  
\begin{equation}
    \text{Count}_{\text{sector}}(i,j)
    = \Bigl|\bigl\{\, u \in [t-L,t-1] \colon r_{u,S(i)}^{\text{sector}} > r_{u,S(j)}^{\text{sector}}\bigr\}\Bigr|.    
\end{equation}

To encode these pairwise relationships into a form suitable for LLM-based embeddings, we generate textual prompts that synthesize the computed statistics. Let $p_{i,j}$ be a prompt-generating function that takes as input the historical returns $\{r_{t-L:t-1,i}, r_{t-L:t-1,j}\}$, sector assignments $(S(i), S(j))$, sector-level yields $\{r_{u,S(i)}^{\text{sector}}\}_{u=t-L}^{t-1}$, $\{r_{u,S(j)}^{\text{sector}}\}_{u=t-L}^{t-1}$, and the pairwise performance statistics $\text{Count}_{\text{stock}}(i,j)$ and $\text{Count}_{\text{sector}}(i,j)$. This function produces a textual prompt describing the relative performance and sectoral context of the two assets.

Collecting such prompts for all pairs $(i,j)$ with $i \neq j$ yields:
\begin{equation}
        \mathcal{P}_{\text{Stocks}} 
    = \bigl\{\, p_{i,j}\bigl([\text{Count}_{\text{stock}}(i,j),\, \text{Count}_{\text{sector}}(i,j)]\bigr) 
      : i \neq j \bigr\},
\end{equation}
where $[\text{a},\text{b}]$ denotes the concatenation of inputs into a single composite prompt for $p_{i,j}$. The prompt-generation function $p_{i,j}$ is a function that maps $\mathcal{T}$ to the space of text descriptions.

Each prompt in $\mathcal{P}_{\text{Stocks}}$ is mapped to a token-level representation via the LLM embedding function $g_{\phi}(\cdot)$. We then stack or concatenate these token embeddings across all prompts, yielding
\begin{equation}
 E_{\text{stocks}} = \bigl\{g_{\phi}(p)\bigr\}_{p \in \mathcal{P}_{\text{Stocks}}} 
     \in \mathbb{R}^{M_{\text{stocks}} \times d_{\text{LLM}}},
\end{equation}
where $M_{\text{stocks}}$ represents the total token count across all stock-related prompts. This embedding $E_{\text{stocks}}$ encodes both asset-level relationships, drawn from pairwise performance statistics, and sector-level relationships, informed by aggregated sector yields and asset-to-sector mappings.

\textbf{Macroeconomic embeddings:} While the above embeddings capture asset-level interactions and sectoral dynamics, they do not fully account for the broader macroeconomic environment. Macroeconomic factors often shape market conditions, influencing correlations among assets and risk-return profiles. However, macroeconomic indicators are frequently observed at irregular intervals and may not align with the regular sampling of financial returns. Directly integrating these irregular observations can pose significant technical and modeling challenges.
To address this, we map macroeconomic data into textual descriptions that summarize their key characteristics. Let $x_{t} \in \mathbb{R}^{M}$ denote a vector of $M$ macroeconomic variables observed at time $t$. Since not all indicators are observed at every time point, let $\mathcal{T}_{m}=\{t_{1}^{(m)}, t_{2}^{(m)},\ldots,t_{|\mathcal{T}_{m}|}^{(m)}\}$ be the set of observation times for the $m$-th variable.

\begin{figure}[h!] 
  \centering
  \includegraphics[width=\linewidth]{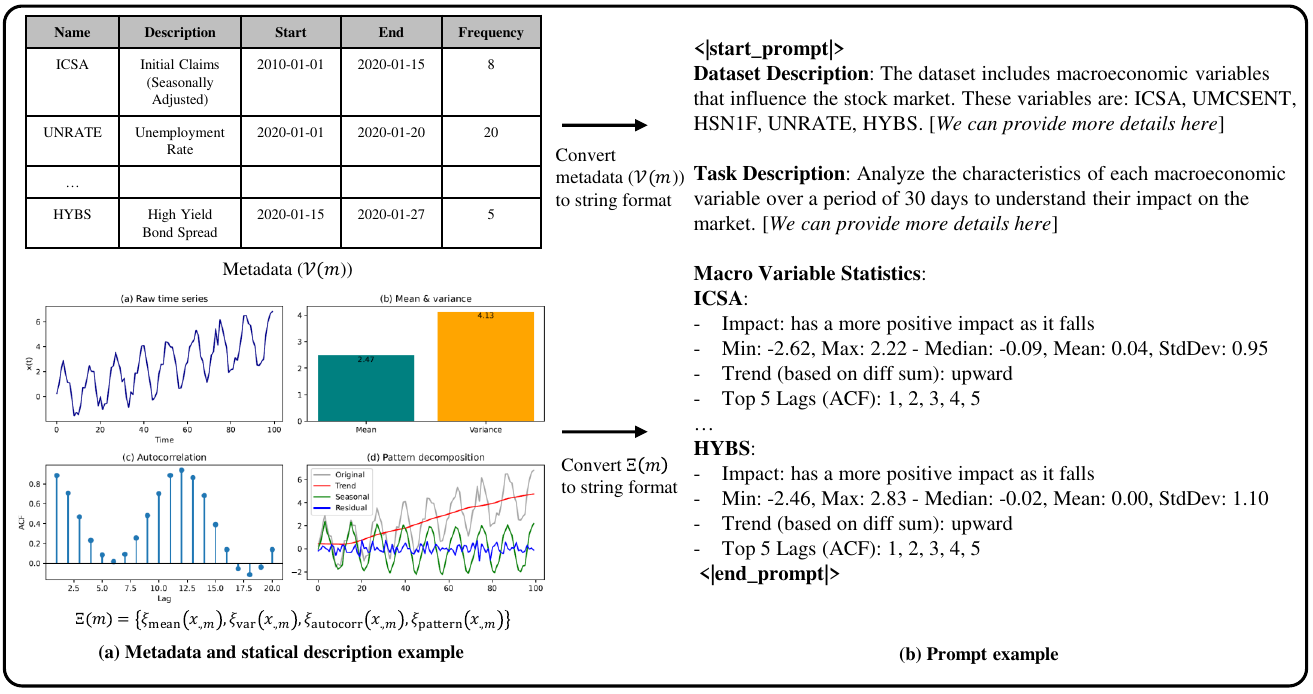}
   \captionsetup{font=footnotesize}
   \caption{Illustration of LLM-based prompt generation from pairwise outperformance statistics and macroeconomic summaries. For each asset pair, relative performance and sector-level yields are synthesized into textual prompts that capture inter-asset relationships, while similarly constructed macro-level prompts summarize irregularly observed economic indicators. These textual prompts are embedded by the LLM and subsequently integrated, via the cross-attention mechanism, into the DINN architecture.}
  \label{fig:prompt}
\end{figure}

Following \citep{jin2023time}, we extend this approach to handle irregularly sampled variables explicitly. Define a set of transformations $\Xi=\{\xi_{\text{mean}},\xi_{\text{var}},\xi_{\text{autocorr}},\xi_{\text{pattern}}\}$, each capable of extracting salient features from the irregularly sampled observations $\{x_{t,m} : t \in \mathcal{T}_{m}\}$: 
\begin{equation}
    \Xi(m)=\{\xi_{\text{mean}}(x_{\cdot,m}),\,\xi_{\text{var}}(x_{\cdot,m}),\,\xi_{\text{autocorr}}(x_{\cdot,m}),\,\xi_{\text{pattern}}(x_{\cdot,m})\}
\end{equation}
where $x_{\cdot,m}$ denotes all observed values of the $m$-th macroeconomic variable. Each $\xi_{\cdot}$ operator is defined to accommodate irregular time intervals, ensuring accurate representation of the underlying statistical properties. An illustration of this prompt-generation and embedding process for macroeconomic features is shown in \Cref{fig:prompt}.

Now, for each variable $m$, let $q_{m}$ be a prompt-generating function $q_{m}: \left(\{x_{t,m}\}_{t \in \mathcal{T}_{m}},\mathcal{V}(m)\right) \to \mathcal{T}$ where $\mathcal{V}(m)$ denotes any auxiliary metadata for variable $m$, and $\mathcal{T}$ is the space of textual descriptions. The function $q_{m}$ synthesizes the extracted statistics $\Xi(m)$ and metadata $\mathcal{V}(m)$ into a coherent textual summary. This textual prompt could, for example, note that a given macroeconomic variable has been trending upward, showing seasonal patterns or strong autocorrelation. The function $q_{m}$ synthesizes the extracted statistics $\Xi(m)$ and metadata $\mathcal{V}(m)$ into a coherent textual summary. This textual prompt could, for example, note that a given macroeconomic variable has been trending upward, showing seasonal patterns or strong autocorrelation. Collecting these prompts across all $M$ macroeconomic variables:
\begin{equation} 
    \mathcal{P}_{\text{Macro}} = \{q_{m}(\{x_{t,m}\}_{t\in \mathcal{T}_{m}}, \mathcal{V}(m)) : m=1,\ldots,M \}
\end{equation}

Let $g_{\phi}(\cdot)$ be the same pretrained LLM embedding function used for the inter-asset relationship embeddings. Applying it to each prompt in $\mathcal{P}_{\text{Macro}}$ yields a sequence of token-level representations, which we then stack to form

\begin{equation}
    \mathbf{E}_{\text{macro}} 
    = \bigl\{\,g_{\phi}(p)\bigr\}_{p \in \mathcal{P}_{\text{Macro}}}
    \;\in\;\mathbb{R}^{M_{\text{macro}} \times d_{\text{LLM}}},    
\end{equation}

where $M_{\text{macro}}$ denotes the total token count across all macro prompts. The resulting embedding, $\mathbf{E}_{\text{macro}}$, captures broader economic context complementary to the asset-level embeddings. By reflecting trends, volatility, and structural patterns of macroeconomic variables through natural-language prompts, it enriches the overall representational scope of the model. Unlike previous methods relying solely on return-based factor structures extracted from asset movements \citep{zhang2021universal, giglio2022factor, chen2024deep}, our approach integrates macroeconomic context through semantic embeddings grounded in LLMs.

\subsection{Decision-informed neural network} In this section, we present our neural network that integrates multi-modal information for portfolio optimization. The architecture consists of four key components: (1) a cross-attention mechanism that fuses temporal patterns with LLM-derived semantic embeddings, (2) a pretrained large language model for return forecasts, (3) a differentiable optimization layer that converts predictions into portfolio weights, and (4) a hybrid training objective combining forecasting and decision-focused losses. 

\subsubsection{Efficient Dual-Modality Integration via Prob-Sparse Cross-Attention} Given the decomposed normalized returns and LLM-based embeddings, we employ a prob-sparse cross-attention mechanism \citep{informer_2021} to integrate temporal and semantic information efficiently. In a naive full attention framework \citep{waswani2017attention}, the computational cost scales proportionally to the product of query and key lengths in the simplest case), which becomes prohibitively large for long sequences of textual embeddings or when $N$ and $M$ grow significantly. By contrast, prob‐sparse attention uses a sampling‐based approximation that retains only the most relevant keys for each query. Specifically, for each query, it selects a subset of key positions whose dot‐products are likely to dominate the attention distribution, thereby reducing the effective number of terms in the softmax normalization. This approach substantially lowers the computational complexity under common parameter choices), while preserving the representational capacity and accuracy of attention‐based models.

We employ prob‐sparse attention for two main reasons. First, it alleviates computational and memory burdens that arise from large collections of textual or macroeconomic embeddings, ensuring scalability for real‐world financial datasets with many assets and extended textual descriptions. Second, this approximation focuses model capacity on salient interactions, often leading to improved efficiency during training without sacrificing forecast fidelity. 

Let $\mathbf{T}_{t} \in \mathbb{R}^{L \times N}$ and $\mathbf{R}_{t} \in \mathbb{R}^{L \times N}$ denote the trend and residual components respectively, where $ \mathbf{T}_{t} = [\tau_{t-L+1},\tau_{t-L+2},\ldots,\tau_{t-1}]^\top$ and $ \mathbf{R}_{t} = [\rho_{t-L+1},\rho_{t-L+2},\ldots,\rho_{t-1}]^\top$. The LLM-based semantic embeddings are represented as $\mathbf{E}_{\text{stocks}} \in \mathbb{R}^{M_{\text{stocks}} \times d_{\text{LLM}}}$ and $\mathbf{E}_{\text{macro}} \in \mathbb{R}^{M_{\text{macro}} \times d_{\text{LLM}}}$, where $d_{\text{LLM}}$ denotes the embedding dimension, and $M_{\text{stocks}}, M_{\text{macro}}$ represent the respective sequence lengths of the textual embeddings. We define a cross-attention operation $\text{CrossAttn}(\mathbf{X}, \mathbf{Y})$ that maps temporal patterns $\mathbf{X} \in \mathbb{R}^{L \times N}$ and textual embeddings $\mathbf{Y} \in \mathbb{R}^{M \times d_{\text{LLM}}}$ into an integrated representation in $\mathbb{R}^{N \times d_{\text{LLM}}}$. First, we transpose the temporal input to $\mathbf{X}^\prime = \mathbf{X}^\top \in \mathbb{R}^{N \times L}$ to align the asset dimension with the attention mechanism. Next, we compute query, key, and value representations through learnable linear transformations:

\begin{equation}
    \mathbf{Q} = \mathbf{X}^\prime \mathbf{W}^Q, \quad \mathbf{K} = \mathbf{Y} \mathbf{W}^K, \quad \mathbf{V} = \mathbf{Y} \mathbf{W}^V,
\end{equation}

where $\mathbf{W}^Q, \mathbf{W}^K, \mathbf{W}^V \in \mathbb{R}^{d_{\text{LLM}} \times d_{\text{LLM}}}$ are learnable parameters. To enhance representational capacity, we employ multi-head attention with $B$ heads, each of dimension $d_b$ such that $B \times d_b = d_{\text{LLM}}$. The matrices $\mathbf{Q}, \mathbf{K}, \mathbf{V}$ are split across heads:

\begin{equation}
    \mathbf{Q} \to [\mathbf{Q}_1,\ldots,\mathbf{Q}_B], \quad \mathbf{K} \to [\mathbf{K}_1,\ldots,\mathbf{K}_B], \quad \mathbf{V} \to [\mathbf{V}_1,\ldots,\mathbf{V}_B],
\end{equation}

where $\mathbf{Q}_b \in \mathbb{R}^{N \times d_b}$ and $\mathbf{K}_b, \mathbf{V}_b \in \mathbb{R}^{M \times d_b}$ for each head $b \in \{1,\ldots,B\}$.

Following the prob-sparse attention mechanism \citep{informer_2021}, we compute a sparse approximation of the attention weights. Let $c>0$ be a constant and define $U_b = c\lceil \log M \rceil$ as the number of sampled key positions. For each query position $i \in \{1,\ldots,N\}$, we sample a subset $\mathcal{S}_b(i) \subseteq \{1,\ldots,M\}$ of size $U_b$. The attention weights for head $b$ are:
\begin{equation}
    \alpha_{b,i,j} = 
    \begin{cases}
        \frac{\exp\left(\frac{(\mathbf{Q}_b)_{i,:}(\mathbf{K}_b)_{j,:}^\top}{\sqrt{d_b}}\right)}{\sum_{j' \in \mathcal{S}_b(i)} \exp\left(\frac{(\mathbf{Q}_b)_{i,:}(\mathbf{K}_b)_{j',:}^\top}{\sqrt{d_b}}\right)} & \text{if } j \in \mathcal{S}_b(i), \\
        0 & \text{otherwise}.
    \end{cases}
\end{equation}

So, the output for each head is computed as:
\begin{equation}
    (\mathbf{Z}_b)_{i,:} = \sum_{j \in \mathcal{S}_b(i)} \alpha_{b,i,j} (\mathbf{V}_b)_{j,:},
\end{equation}

and the final output is obtained by concatenating across heads and applying a linear projection:

\begin{equation}
    \mathbf{Z} = [\mathbf{Z}_1;\ldots;\mathbf{Z}_B]\mathbf{W}^O \in \mathbb{R}^{N \times d_{\text{LLM}}},
\end{equation}

where $\mathbf{W}^O \in \mathbb{R}^{(B d_b) \times d_{\text{LLM}}}$ is a learnable parameter matrix. Then, we apply this cross-attention mechanism separately to integrate market-level and stock-specific information: 
\begin{equation}
\begin{aligned}
    \mathbf{C}_{\text{market}} &= \text{CrossAttn}(\mathbf{T}_t, \mathbf{E}_{\text{macro}}) \in \mathbb{R}^{N \times d_{\text{LLM}}}, \\
    \mathbf{C}_{\text{stock}} &= \text{CrossAttn}(\mathbf{R}_t, \mathbf{E}_{\text{stocks}}) \in \mathbb{R}^{N \times d_{\text{LLM}}}.
\end{aligned}
\end{equation}

The resulting representations $\mathbf{C}_{\text{market}}$ and $\mathbf{C}_{\text{stock}}$ capture the alignment between temporal patterns and semantic embeddings at both market and individual stock levels. This dual representation in a common $d_{\text{LLM}}$-dimensional space can facilitates the subsequent joint modeling of returns and portfolio optimization.

\subsubsection{Pretrained large language model for prediction} With the integrated representations from the cross-attention mechanism, we leverage a pretrained large language model to generate return forecasts. Let $g_{\phi}: \mathbb{R}^{N \times d_{\text{LLM}}} \to \mathbb{R}^{N \times d_{\text{LLM}}}$ be the pretrained LLM with frozen parameters $\phi$. It serves as a fixed contextual encoder that maps integrated embeddings into a more semantically enriched space.
Given $\mathbf{C}_{\text{market}}, \mathbf{C}_{\text{stock}} \in \mathbb{R}^{N \times d_{\text{LLM}}}$, we process them through the LLM:
\begin{equation}
\mathbf{Z}_{\text{market}} = g_{\phi}(\mathbf{C}_{\text{market}}), \quad
\mathbf{Z}_{\text{stock}} = g_{\phi}(\mathbf{C}_{\text{stock}})
\end{equation}
where $\mathbf{Z}_{\text{market}}, \mathbf{Z}_{\text{stock}} \in \mathbb{R}^{N \times d_{\text{LLM}}}$. The LLM refines these embeddings by capturing higher-order dependencies among assets through its attention mechanisms while preserving the semantic information encoded in the original representations.

To combine the market-level and stock-specific information, we employ an additive fusion $\mathbf{Z} = \mathbf{Z}_{\text{market}} + \mathbf{Z}_{\text{stock}} \in \mathbb{R}^{N \times d_{\text{LLM}}}$, where the addition is performed element-wise. This operation assumes both embeddings reside in a common semantic space and that their contributions to the final representation are complementary.

To generate normalized return forecasts over the horizon $H$, we project the fused embeddings through a learned linear transformation $\hat{r}^{\prime}_{t+1:t+H} = (\mathbf{Z} \mathbf{W}^{F})^{\top}$, where $\mathbf{W}^{F} \in \mathbb{R}^{d_{\text{LLM}} \times H}$ is a trainable weight matrix and $\hat{r}^{\prime}_{t+1:t+H} \in \mathbb{R}^{H \times N}$. To recover the returns in their original scale, we apply the inverse of the normalization transformation introduced in Section 3.2.1. For each asset $i$ and horizon $h$, we denormalize the predictions using the historical statistics:
\begin{equation}
\hat{r}_{t+h,i} = \hat{r}^{\prime}_{t+h,i} \sigma_{t,i} + \mu_{t,i}
\end{equation}
where $\mu_{t,i}$ and $\sigma_{t,i}$ are the sample mean and standard deviation computed over the lookback window $[t-L, t-1]$ as defined previously.

The final return predictions can be organized into a matrix $\hat{r}_{t+1:t+H} = [\hat{r}_{t+1}, \hat{r}_{t+2}, \ldots, \hat{r}_{t+H}] \in \mathbb{R}^{H \times N}$, where each $\hat{r}_{t+h} \in \mathbb{R}^{N}$ represents the predicted returns across all assets at time $t+h$. 

While employing the latest pretrained LLMs can significantly boost predictive performance, it also raises a critical concern of \emph{data leakage} in empirical evaluations. Because some LLMs (e.g., GPT-4o \citep{achiam2023gpt}, LLAMA \citep{dubey2024llama}) were trained on vast text corpora---potentially including financial data, news reports, or research materials overlapping with one’s test set---there is a nontrivial risk that information from the true “future” may already reside within the LLM’s parameters. Consequently, evaluating forecasts on a test period that the LLM might have indirectly “seen” during pretraining can yield overly optimistic results. Therefore, we used the GPT-2, which is a relatively old model with sufficient representation power, as the default LLM model to avoid the issue of data leakage.

\subsubsection{Optimization layer} \label{optim_layer} The optimization layer converts predicted returns into optimal portfolio weights by solving a convex optimization problem that balances expected returns and portfolio risk. Given predicted returns $\hat{r}_{t+1:t+H} \in \mathbb{R}^{H \times N}$ and historical returns $r_{t-K:t-1}^{\star} \in \mathbb{R}^{K \times N}$, we estimate covariance matrices $\hat{\Sigma}_{t+h}$ by combining historical and predicted return as $\hat{\Sigma}_{t+h} = \text{Cov}\left( r_{t-K:t-1}^{\star} \cup \hat{r}_{t+1:t+h} \right)$. In this study, we use the past three months of historical returns for stable covariance estimation. Assuming $\hat{\Sigma}_{t+h}$ is positive definite, we perform a Cholesky decomposition $\hat{\Sigma}_{t+h} = \hat{L}_{t+h}\hat{L}_{t+h}^{\top}$. Let $\lambda>0$ be the risk-aversion parameter. For each time step $t+h$, we solve:

\begin{equation}
\begin{aligned}
\min_{w_{t+h}} \quad & \lambda s_{t+h}^2 - \hat{\mu}_{t+h}^{\top} w_{t+h} \\
\text{s.t.} \quad & \|\hat{L}_{t+h} w_{t+h}\|_2 \leq s_{t+h}, \\
                      & s_{t+h} \geq 0, \\
                      & \sum_{i=1}^{N} w_{t+h,i} = 1, \\
                      & 0 \leq w_{t+h,i} \leq 1 \quad \forall i \in \{1, \ldots, N\}.
\end{aligned}
\end{equation}

Here, $s_{t+h}$ represents the portfolio volatility, and the full-investment, long-only constraints ensure that $\sum_i w_{t+h,i} = 1$ with $w_{t+h,i} \ge 0$. This second-order cone formulation is equivalent to solving a mean–variance trade-off problem, where $\lambda$ modulates the level of risk-aversion, and $\hat{\mu}_{t+h}$ encodes the expected return estimates.  Solving this second-order cone optimization problem for each $h$ yields:
\begin{equation}
    \hat{w}_{t+1:t+H} = \bigl(\hat{w}_{t+1}, \hat{w}_{t+2}, \ldots, \hat{w}_{t+H}\bigr)
\end{equation}

\subsubsection{Training} Training aims to align the model’s parameters so that the predicted returns and the resulting decision-making process closely approximate their true counterparts. To achieve this, we combine a forecasting loss and a decision-focused loss into a single objective function. Let $\hat{r}_{t:t+H}$ be the predicted returns over the horizon $H$, and $r_{t:t+H}$ be the corresponding actual returns. The first loss term, which we denote as the forecasting loss, is the mean squared error (MSE) computed over the forecast horizon:
\begin{equation}
 \mathcal{L}_{\mathrm{MSE}}=\frac{1}{NH} \sum_{h=1}^H\left\|\hat{r}_{t+h}-r_{t+h}\right\|_2^2   
\end{equation}
The decision-focused loss measures how prediction errors degrade portfolio quality. Consider optimal weights $w_{t+1:t+h}^{\star}$ obtained from actual returns and $\hat{w}_{t+1:t+h}$ from predicted returns. With $L_{t+h}^{\star}$ the Cholesky factor of the actual covariance $\Sigma_{t+h}^{\star}$, define:
\begin{equation}
    \begin{gathered}
J_{t+h}^{\star}=\lambda\left\|L_{t+h}^{\star} w_{t+h}^{\star}\right\|_2-\mu_{t+h}^{\star \top} w_{t+h}^{\star}, \\
\hat{J}_{t+h}=\lambda\left\|L_{t+h}^{\star} \hat{w}_{t+h}\right\|_2-\mu_{t+h}^{\star \top} \hat{w}_{t+h} 
\end{gathered}
\label{eq:13}
\end{equation}

Intuitively, these performance measures quantify how inaccuracies in predicted returns translate into suboptimal portfolio decisions. Unlike approaches such as those in \citep{costa2023distributionally}, which optimize for metrics like the Sharpe ratio, the proposed decision-focused loss directly measures the regret incurred by substituting predicted returns for actual ones. Consequently, $\Delta J_{t+h} = \hat{J}_{t+h}-J_{t+h}^{\star}$ reflects the additional cost induced by prediction errors on the portfolio’s true risk-return profile. Then the decision-focused loss is the average absolute regret as here:
\begin{equation}
     \mathcal{L}_{\mathrm{Decision}}=\frac{1}{NH} \sum_{h=1}^H|\Delta J_{t+h}|.
\end{equation}
where $\Delta J_{t+h}$ is the discrepancy between the performance of the predicted and true portfolios. 

Finally, we combine the two losses into a single training objective using a weighting parameter $\beta \in [0,1]$, which balances between predictive accuracy and decision robustness:
\begin{equation}
    \mathcal{L}_{\mathrm{loss}}=\beta \mathcal{L}_{\mathrm{MSE}}+(1-\beta) \mathcal{L}_{\mathrm{Decision}}
\end{equation}
By adjusting $\beta$, we can control the relative importance of minimizing forecast errors versus minimizing decision regret. We set $\beta$ = 0.4 as the default value in this study. 

\subsection{Gradient for optimization problem} Consider the decision-focused loss $\mathcal{L}_{\mathrm{Decision}}$, which measures how predictive inaccuracies translate into suboptimal portfolio choices. This loss depends on the model parameters $\theta$ through the predicted returns. Since the predicted returns determine $\hat{\mu}_{t+h}$ and $\hat{L}_{t+h}$, the optimal weights $\hat{w}_{t+h}$ obtained from the optimization layer also depend implicitly on $\theta$.

Define $\Delta J_{t+h} = \hat{J}_{t+h} - J_{t+h}^{\star}$, where $J_{t+h}^{\star} = \lambda \|L_{t+h}^{\star} w_{t+h}^{\star}\|_2 - \mu_{t+h}^{\star \top} w_{t+h}^{\star}$ is the benchmark performance using true returns and true covariance, and $\hat{J}_{t+h} = \lambda \|L_{t+h}^{\star} \hat{w}_{t+h}\|_2 - \mu_{t+h}^{\star \top} \hat{w}_{t+h}$ is the performance under predicted quantities and weights. Since $J_{t+h}^{\star}$ does not depend on $\theta$, its gradient is zero. Thus, the gradient of $\mathcal{L}_{\mathrm{Decision}}$ with respect to $\theta$ reduces to the gradient of $\hat{J}_{t+h}$.

Ignoring non-differentiability at zero for the absolute value and assuming a differentiable approximation if needed, the derivative of $\hat{J}_{t+h}$ with respect to $\hat{w}_{t+h}$ is

\begin{equation}
\begin{aligned}
\nabla_{\hat{w}_{t+h}}\hat{J}_{t+h} &= \nabla_{\hat{w}_{t+h}}\left(\lambda \|L_{t+h}^{\star}\hat{w}_{t+h}\|_2 - \mu_{t+h}^{\star \top}\hat{w}_{t+h}\right) \\
&= \nabla_{\hat{w}_{t+h}}\left(\lambda \|L_{t+h}^{\star}\hat{w}_{t+h}\|_2\right) 
- \nabla_{\hat{w}_{t+h}}\left(\mu_{t+h}^{\star \top}\hat{w}_{t+h}\right) \\
&= \lambda \nabla_{\hat{w}_{t+h}}\sqrt{(L_{t+h}^{\star}\hat{w}_{t+h})^{\top}(L_{t+h}^{\star}\hat{w}_{t+h})} 
- \mu_{t+h}^{\star} \\
&= \lambda \frac{L_{t+h}^{\star \top}(L_{t+h}^{\star}\hat{w}_{t+h})}{\sqrt{(L_{t+h}^{\star}\hat{w}_{t+h})^{\top}(L_{t+h}^{\star}\hat{w}_{t+h})}} 
- \mu_{t+h}^{\star} \\
&= \lambda \frac{L_{t+h}^{\star \top}(L_{t+h}^{\star}\hat{w}_{t+h})}{\|L_{t+h}^{\star}\hat{w}_{t+h}\|_2} - \mu_{t+h}^{\star}.
\end{aligned}
\end{equation}

This gradient provides the directional sensitivity of the performance measure $\hat{J}_{t+h}$ to changes in the predicted weights.

Because $\hat{w}_{t+h}$ solves a parametric optimization problem whose parameters $\hat{\mu}_{t+h}$ and $\hat{L}_{t+h}$ depend on $\theta$, the chain rule must be applied to propagate gradients through the optimization layer. Formally, let $\mathcal{L}_{\mathrm{Decision}}$ be defined as an average over the forecast horizon:

\begin{equation}
    \nabla_{\theta}\mathcal{L}_{\mathrm{Decision}} = \frac{1}{NH}\sum_{h=1}^{H} \left( \frac{\partial \mathcal{L}_{\mathrm{Decision}}}{\partial \hat{w}_{t+h}}\frac{\partial \hat{w}_{t+h}}{\partial \hat{\mu}_{t+h}}\frac{\partial \hat{\mu}_{t+h}}{\partial \theta} + \frac{\partial \mathcal{L}_{\mathrm{Decision}}}{\partial \hat{w}_{t+h}}\frac{\partial \hat{w}_{t+h}}{\partial \hat{L}_{t+h}}\frac{\partial \hat{L}_{t+h}}{\partial \theta}\right)
\end{equation}

Here, $\partial \hat{w}_{t+h}/\partial \hat{\mu}_{t+h}$ and $\partial \hat{w}_{t+h}/\partial \hat{L}_{t+h}$ quantify the sensitivities of the optimal weights to perturbations in predicted means and covariance factors, respectively. These can be derived via the implicit function theorem or through established results in parametric optimization. The terms $\partial \hat{\mu}_{t+h}/\partial \theta$ and $\partial \hat{L}_{t+h}/\partial \theta$ capture how the predictive model’s parameters $\theta$ affect the predicted inputs to the optimization layer.However, While computing the sensitivity terms $\partial \hat{w}_{t+h}/\partial \hat{\mu}_{t+h}$ and $\partial \hat{w}_{t+h}/\partial \hat{L}_{t+h}$ is computationally challenging due to the implicit nature of the optimization problem's solution, these derivatives provide valuable information about how estimation errors in predicted moments affect optimal portfolio weights. As demonstrated in Theorems 1 and Theorems 2, under appropriate regularity conditions, these sensitivities can be characterized using the implicit function theorem applied to the KKT conditions, enabling efficient gradient-based learning through the optimization layer.

\textbf{Theorem 1 (Sensitivity of optimal portfolio weights w.r.t. predicted returns)} Consider the following portfolio optimization problem at each time step $ t+h $ for $\{h = {1, \cdots, H}\}$:
\begin{equation}
\begin{aligned}
\min_{\hat{w}_{t+h}} \quad & \lambda s_{t+h}^{2} - \hat{\mu}_{t+h}^{\top} \hat{w}_{t+h} \\
\text{subject to} \quad & \|\hat{L}_{t+h} \hat{w}_{t+h}\|_2 = s_{t+h}, \\
& \sum_{i=1}^{N} \hat{w}_{t+h,i} = 1,
\end{aligned}
\label{eq:modified_problem}
\end{equation}

where $\lambda > 0$, $\hat{w}_{t+h} \in \mathbb{R}^N$ denotes the portfolio weights, $\hat{\mu}_{t+h} \in \mathbb{R}^N$ are the predicted returns, and $\hat{L}_{t+h} \in \mathbb{R}^{N \times N}$ is a lower-triangular Cholesky factor such that $\hat{\Sigma}_{t+h} = \hat{L}_{t+h}\hat{L}_{t+h}^{\top}$ is the covariance matrix of returns. Assume $\hat{\Sigma}_{t+h}$ is invertible.

Then the derivative of the optimal solution $\hat{w}_{t+h}$ with respect to the predicted returns $\hat{\mu}_{t+h}$ is given by:
\begin{equation}
    \frac{\partial \hat{w}_{t+h}}{\partial \hat{\mu}_{t+h}} = \hat{\Sigma}_{t+h}^{-1} \;-\; \frac{\hat{\Sigma}_{t+h}^{-1}\mathbf{1}\mathbf{1}^{\top}\hat{\Sigma}_{t+h}^{-1}}{\mathbf{1}^{\top}\hat{\Sigma}_{t+h}^{-1}\mathbf{1}},
\end{equation}
where $\mathbf{1}$ is an $N$-dimensional vector of ones.

\textbf{Proof: } Refer to Appendix A.1 for a detailed derivation, which follows from applying Lagrangian duality and differentiating the resulting Karush-Kuhn-Tucker (KKT) conditions with respect to $\hat{\mu}_{t+h}$.

\textbf{Theorem 2 (Sensitivity of optimal portfolio weights w.r.t. cholesky factor)}
Under the same setting and assumptions as in Theorem 1, let
\begin{equation}
    p := \frac{\mathbf{1}^{\top}\hat{\Sigma}_{t+h}^{-1}\hat{\mu}_{t+h} - 1}{\mathbf{1}^{\top}\hat{\Sigma}_{t+h}^{-1}\mathbf{1}}
\end{equation}

Then the derivative of the optimal solution $\hat{w}_{t+h}$ with respect to the cholesky factor $\hat{L}_{t+h}$ is given by:
\begin{equation}
    \frac{\partial \hat{w}_{t+h}}{\partial \hat{L}_{t+h}} = -2\, \hat{\Sigma}_{t+h}^{-1}(\hat{\mu}_{t+h} - p\mathbf{1})\hat{\Sigma}_{t+h}^{-1}\hat{L}_{t+h} \;-\; 2\, \hat{\Sigma}_{t+h}^{-1}\left( \frac{\partial p}{\partial \hat{\Sigma}_{t+h}}\mathbf{1} \right)\hat{L}_{t+h},
\end{equation}

where $\frac{\partial p}{\partial \hat{\Sigma}_{t+h}} = \frac{- \hat{\Sigma}_{t+h}^{-1}\mathbf{1}\hat{\mu}_{t+h}^{\top}\hat{\Sigma}_{t+h}^{-1} z \;+\; (\mathbf{1}^{\top}\hat{\Sigma}_{t+h}^{-1}\hat{\mu}_{t+h} - 1)\hat{\Sigma}_{t+h}^{-1}\mathbf{1}\mathbf{1}^{\top}\hat{\Sigma}_{t+h}^{-1}}{z^{2}}
$ and $z = \mathbf{1}^{\top}\hat{\Sigma}_{t+h}^{-1}\mathbf{1}.$

\textbf{Proof: } Refer to Appendix A.1 for a detailed proof, which follows by applying the chain rule to the Markowitz optimization problem and carefully differentiating with respect to the Cholesky factor $\hat{L}_{t+h}$. These expressions provide explicit formulas for the sensitivities needed to efficiently implement gradient-based learning through the optimization layer, enabling a deeper understanding of how inaccuracies in predicted inputs influence optimal decision-making.

\label{ch:ch4}
\section{Experiment}
We now present the experimental results that comprehensively demonstrate the performance of \texttt{DINN} on real-world benchmark datasets. To facilitate transparency and reproducibility, the code and configuration details are available at \href{https://anonymous.4open.science/r/Decision-informed-Neural-Networks-with-Large-Language-Model-Integration-for-Portfolio-Optimization-A441/README.md}{Anonymous Github}.

\subsection{Implementation details}
In this section, we describe the datasets, evaluation metrics, baseline models, and hyperparameter settings used in our empirical study. 

\subsubsection{Dataset description}
This study analyzes a comprehensive dataset spanning January 2010 to December 2023, encompassing both the post-financial crisis recovery and the COVID-19 pandemic period. Our primary data consists of equity returns from two major indices: the DOW 30 and a market-cap-weighted subset of 50 constituents from the S\&P 100. To address potential survivorship bias, we include only companies that maintained consistent index membership throughout the study period. The financial data, obtained from WRDS, is complemented by five macroeconomic indicators from FRED, selected based on their documented predictive power in asset pricing: weekly initial jobless claims (ICSA), consumer sentiment (UMCSENT), new home sales (HSN1F), unemployment rate (UNRATE), and high-yield bond spread (HYBS). These variables may capture different aspects of economic conditions that influence asset returns through both systematic risk channels and behavioral mechanisms.

\subsubsection{Evaluation Metrics}
We evaluate each model using eight key metrics designed to capture both return characteristics and various dimensions of risk. These include:

\begin{enumerate}
    \item \textbf{Annualized Return (Ret)}: Reflects the average annual growth of the portfolio without subtracting any risk-free component.
    \item \textbf{Annualized Standard Deviation (Std)}: Gauges the volatility of returns, serving as a basic measure of risk.
    \item \textbf{Sharpe Ratio (SR)}: Examines excess returns (portfolio return minus the risk-free rate) per unit of total volatility.
    \item \textbf{Sortino Ratio (SOR)}: Focuses on downside volatility, isolating harmful fluctuations from benign ones.
    \item \textbf{Maximum Drawdown (MDD)}: Captures the largest observed loss from a prior portfolio high, providing a measure of potential capital erosion.
    \item \textbf{Value at Risk (VaR) at 95\% (monthly)}: Indicates the worst likely loss over a specific time horizon under normal market conditions.
    \item \textbf{Return Over VaR (RoV)}: Scales the portfolio’s excess monthly returns relative to VaR, highlighting returns per tail-risk unit.
    \item \textbf{Terminal Wealth (Wealth)}: Reflects the final cumulative portfolio value, integrating the impact of both returns and compounding.
\end{enumerate}

\subsubsection{Baseline Models and Hyperparameter Selection}
We compare \texttt{DINN} against several state-of-the-art deep learning architectures tailored to financial time series, including both Transformer-based and large language model (LLM)-based methods:

\begin{itemize}
    \item \textbf{Transformer-based methods}: iTransformer \citep{liu2023itransformer}, PatchTST \citep{Yuqietal_PatchTST}, TimesNet \citep{wu2023timesnet}, and Crossformer \citep{zhang2023crossformer}.
    \item \textbf{LLM-based methods}: PAttn \citep{tan2024language}, Chronos \citep{ansari2024chronos}, and GPT4TS \citep{zhou2023one}.
\end{itemize}

All baseline models are implemented using their original architectures and recommended hyperparameters, with minor refinements to accommodate the specifics of our financial data. We provide the detailed hyperparameter settings for \texttt{DINN} in Appendix A.3., ensuring reproducibility and clarity.

\subsection{Can \texttt{DINN} exceed standard deep learning models for portfolio optimization?}
Standard deep learning approaches often focus on minimizing forecasting error without directly addressing the inherent fragility of portfolio selection when faced with small parameter estimation errors. Accordingly, even substantial gains in predictive accuracy may not translate into robust improvements in actual investment outcomes. By contrast, \texttt{DINN} integrates portfolio optimization as a learnable module, aligning model parameters not merely to predict returns accurately but also to optimize the final portfolio decision.

\Cref{tab1:main_result} report the eight core performance metrics—annualized return, standard deviation, Sharpe ratio, Sortino ratio, maximum drawdown, Value-at-Risk, Return over VaR, and terminal wealth—across two datasets (S\&P 100 and DOW 30). These indicators each offer a unique perspective on risk and reward.

The empirical evidence demonstrates that \texttt{DINN} consistently outperforms standard deep learning models across multiple dimensions of portfolio performance, particularly in metrics that capture the quality of investment decisions. This superiority manifests in both return generation and risk management, with notably smaller performance variability across experimental trials. In terms of return generation, \texttt{DINN} achieves markedly higher annualized returns of 43.53\% ($\pm$ 1.45\%) for the S\&P 100 and 63.25\% ($\pm$ 0.43\%) for the DOW 30, substantially exceeding the next best performers (TimesNet at 33.28\% $\pm$ 22.54\% and 36.72\% $\pm$ 21.64\%, respectively). More importantly, \texttt{DINN} exhibits relatively low variability in these returns, indicating consistently superior performance rather than sporadic success. This consistency extends to risk-adjusted performance measures, where \texttt{DINN} achieves the highest Sharpe ratios (1.04 $\pm$ 0.04 and 1.29 $\pm$ 0.01) and Sortino ratios (1.50 $\pm$ 0.05 and 1.94 $\pm$ 0.01) across both datasets, again with minimal variability among all models.

The risk management capabilities of \texttt{DINN} reveal a nuanced picture. While GPT4TS achieves marginally lower maximum drawdowns (33.91\% versus \texttt{DINN}'s 39.51\% for S\&P 100), \texttt{DINN} demonstrates remarkably stable risk characteristics, showing the lowest standard deviation in drawdown measures ($\pm$ 0.64\% for S\&P 100, compared to GPT4TS's $\pm$ 3.68\%). This stability is particularly evident in the Value-at-Risk (VaR) metrics, where \texttt{DINN} maintains competitive levels (12.33\% for S\&P 100 and 13.91\% for DOW 30) while exhibiting very small variability ($\pm$ 0.04\% and $\pm$ 0.15\%, respectively).

Most notably, \texttt{DINN} excels in translating its advantages into tangible investment outcomes. The model achieves the highest Return over VaR (19.87\% for S\&P 100 and 27.72\% for DOW 30) and terminal wealth (3.0213 and 4.4715, respectively) for both datasets, with substantially lower variability than competing approaches. This superior wealth accumulation, combined with consistent risk-adjusted performance metrics, suggests that \texttt{DINN}’s decision-informed architecture more effectively bridges the gap between predictive accuracy and portfolio optimization. These empirical results collectively imply that \texttt{DINN} more effectively reconciles predictive accuracy with the practical objectives of portfolio management, yielding robust and reliable performance gains.

\begin{table}[!htbp]
\centering
\centering
{\fontsize{8pt}{11pt}\selectfont
\begin{tabular}{lcccc}
\hline
\multicolumn{5}{l}{\textit{\textbf{Panel A. S\&P 100 Dataset}}}                                                                                                                                                                                                                        \\ 
\multicolumn{1}{c}{\textbf{Measure}} & \textbf{Ret ($\uparrow$)}                                              & \textbf{Std ($\downarrow$)}                                   & \textbf{SR ($\uparrow$)}                          & \textbf{SOR ($\uparrow$)}                         \\ \hline
Crossformer                          & 0.3337 $\pm$ 0.3557                                                      & 0.4529 $\pm$ 0.0789                                             & 0.6468 $\pm$ 0.6428                                 & 1.0090 $\pm$ 1.0336                                 \\
PatchTST                             & 0.0025 $\pm$ 0.0810                                                      & 0.4641 $\pm$ 0.0400                                             & -0.0227 $\pm$ 0.1832                                & -0.0310 $\pm$ 0.2712                                \\
iTransformer                         & 0.2264 $\pm$ 0.3356                                                      & 0.4320 $\pm$ 0.0611                                             & 0.5725 $\pm$ 0.7929                                 & 0.8710 $\pm$ 1.1410                                 \\
TimesNet                             & 0.3328 $\pm$ 0.2254                                                      & 0.3719 $\pm$ 0.0233                                             & 0.8903 $\pm$ 0.6805                                 & 1.2380 $\pm$ 0.9839                                 \\
PAttn                                & -0.0815 $\pm$ 0.0466                                                     & 0.4646 $\pm$ 0.0259                                             & -0.2019 $\pm$ 0.1047                                & -0.3015 $\pm$ 0.1554                                \\
Chronos (Base)                       & 0.1000 $\pm$ 0.1283                                                      & 0.3245 $\pm$ 0.1051                                             & 0.2283 $\pm$ 0.4311                                 & 0.3732 $\pm$ 0.6620                                 \\
Chronos (Large)                      & 0.1608 $\pm$ 0.0976                                                      & \textbf{0.2849} $\pm$ 0.0227                                    & 0.5372 $\pm$ 0.3553                                 & 0.7713 $\pm$ 0.5199                                 \\
GPT4TS                               & 0.2166 $\pm$ 0.0694                                                      & 0.3575 $\pm$ 0.0048                                             & 0.5758 $\pm$ 0.1991                                 & 0.8059 $\pm$ 0.2943                                 \\
DINN (ours)                          & {\color[HTML]{000000} \textbf{0.4353 $\pm$ \color[HTML]{00009B} 0.0145}} & 0.4103 $\pm$ \textbf{\color[HTML]{00009B} 0.0005}                        & \textbf{1.0355 $\pm$ \color[HTML]{00009B} 0.0358}   & \textbf{1.5008 $\pm$ \color[HTML]{00009B} 0.0521}   \\ \hline
\multicolumn{1}{c}{\textbf{Measure}} & \textbf{MDD ($\downarrow$)}                                            & \textbf{VaR ($\downarrow$)}                                   & \textbf{RoV ($\uparrow$)}                         & \textbf{Welath ($\uparrow$)}                      \\ \hline
Crossformer                          & 0.6084 $\pm$ 0.0538                                                      & 0.1602 $\pm$ 0.0336                                             & 0.1109 $\pm$ 0.1029                                 & 1.2204 $\pm$ 0.5301                                 \\
PatchTST                             & 0.7147 $\pm$ 0.0841                                                      & 0.1894 $\pm$ 0.0108                                             & -0.0275 $\pm$ 0.0389                                & {\color[HTML]{000000} 0.6717 $\pm$ 0.1987}          \\
iTransformer                         & 0.5077 $\pm$ 0.2865                                                      & 0.1556 $\pm$ 0.0545                                             & 0.1186 $\pm$ 0.1627                                 & 2.0745 $\pm$ 1.5459                                 \\
TimesNet                             & 0.5107 $\pm$ 0.1052                                                      & 0.1289 $\pm$ 0.0209                                             & 0.1760 $\pm$ 0.1357                                 & 2.7133 $\pm$ 1.9920                                 \\
PAttn                                & 0.7700 $\pm$ 0.0444                                                      & {\color[HTML]{000000} 0.1822 $\pm$ 0.0007}                      & -0.0708 $\pm$ 0.0192                                & 0.4673 $\pm$ \textbf{\color[HTML]{00009B} 0.0863}            \\
Chronos (Base)                       & 0.4365 $\pm$ 0.1739                                                      & 0.1059 $\pm$ 0.0298                                             & 0.0333 $\pm$ 0.1021                                 & 1.2204 $\pm$ 0.5301                                 \\
Chronos (Large)                      & 0.3675 $\pm$ 0.0920                                                      & 0.1131 $\pm$ 0.0321                                             & 0.0915 $\pm$ 0.0814                                 & 1.5877 $\pm$ 0.5051                                 \\
GPT4TS                               & \textbf{0.3391} $\pm$ 0.0368                                             & \textbf{0.1049} $\pm$ 0.0054                                    & 0.1151 $\pm$ 0.0407                                 & 1.7164 $\pm$ 0.3758                                 \\
DINN (ours)                          & 0.3951 $\pm$ \textbf{\color[HTML]{00009B} 0.0064}                               & {\color[HTML]{000000} 0.1233 $\pm$ \textbf{\color[HTML]{00009B} 0.0004}} & \textbf{0.1987 $\pm$ \color[HTML]{00009B} 0.0078}   & \textbf{3.0213} $\pm$ 0.1218                        \\ \hline
\multicolumn{5}{l}{}                                                                                                                                                                                                                                                                  \\ \hline
\multicolumn{5}{l}{\textit{\textbf{Panel B. DOW 30 Dataset}}}                                                                                                                                                                                                                         \\ 
\multicolumn{1}{c}{\textbf{Measure}} & \textbf{Ret ($\uparrow$)}                                              & \textbf{Std ($\downarrow$)}                                   & \textbf{SR ($\uparrow$)}                          & \textbf{SOR ($\uparrow$)}                         \\ \hline
Crossformer                          & 0.1463 $\pm$ 0.2596                                                    & 0.4522 $\pm$ 0.0508                                           & 0.3339 $\pm$ 0.5713                               & 0.5031 $\pm$ 0.8169                               \\
PatchTST                             & 0.1306 $\pm$ 0.0758                                                    & 0.4708 $\pm$ 0.0152                                           & 0.2552 $\pm$ 0.1657                               & 0.4005 $\pm$ 0.2622                               \\
iTransformer                         & 0.1463 $\pm$ 0.2596                                                    & 0.4522 $\pm$ 0.0508                                           & 0.3339 $\pm$ 0.5713                               & 0.5031 $\pm$ 0.8169                               \\
TimesNet                             & 0.3672 $\pm$ 0.2164                                                    & 0.4064 $\pm$ 0.0997                                           & 0.8223 $\pm$ 0.3277                               & 1.1923 $\pm$ 0.6041                               \\
PAttn                                & 0.0692 $\pm$ 0.1522                                                    & 0.4657 $\pm$ 0.0231                                           & 0.1200 $\pm$ 0.3269                               & 0.1897 $\pm$ 0.5028                               \\
Chronos (Base)                       & 0.2364 $\pm$ 0.0989                                                    & 0.3155 $\pm$ 0.0610                                           & 0.7044 $\pm$ 0.2519                               & 1.0094 $\pm$ 0.3411                               \\
Chronos (Large)                      & 0.0598 $\pm$ 0.0676                                                    & \textbf{0.2928} $\pm$ 0.0280                                  & 0.1738 $\pm$ 0.2425                               & 0.2186 $\pm$ 0.3154                               \\
GPT4TS                               & 0.2914 $\pm$ 0.1013                                                    & 0.3579 $\pm$ 0.0049                                           & 0.7827 $\pm$ 0.2820                               & 1.1055 $\pm$ 0.3694                               \\
DINN (ours)                          & \textbf{0.6325 $\pm$ \color[HTML]{00009B} 0.0043}                      & 0.4814 $\pm$ \textbf{\color[HTML]{00009B} 0.0002}                      & \textbf{1.2905 $\pm$ \color[HTML]{00009B} 0.0091} & \textbf{1.9449 $\pm$ \color[HTML]{00009B} 0.0137} \\ \hline
\multicolumn{1}{c}{\textbf{Measure}} & \textbf{MDD ($\downarrow$)}                                            & \textbf{VaR ($\downarrow$)}                                   & \textbf{RoV ($\uparrow$)}                         & \textbf{Welath ($\uparrow$)}                      \\ \hline
Crossformer                          & 0.6617 $\pm$ 0.1405                                                    & 0.1718 $\pm$ 0.0388                                           & 0.0647 $\pm$ 0.1212                               & 1.9403 $\pm$ 0.5746                               \\
PatchTST                             & 0.6236 $\pm$ 0.0891                                                    & 0.1660 $\pm$ 0.0021                                           & 0.0407 $\pm$ 0.0465                               & 1.0770 $\pm$ 0.3151                               \\
iTransformer                         & 0.6617 $\pm$ 0.1405                                                    & 0.1718 $\pm$ 0.0388                                           & 0.0647 $\pm$ 0.1212                               & 1.4170 $\pm$ 1.0613                               \\
TimesNet                             & 0.4742 $\pm$ 0.1498                                                    & 0.1390 $\pm$ 0.0637                                           & 0.1627 $\pm$ 0.0215                               & 2.5758 $\pm$ 1.2002                               \\
PAttn                                & 0.7090 $\pm$ 0.0957                                                    & 0.1673 $\pm$ 0.0002                                           & 0.0036 $\pm$ 0.0764                               & 0.9113 $\pm$ 0.4215                               \\
Chronos (Base)                       & 0.3565 $\pm$ 0.0971                                                    & \textbf{0.0951} $\pm$ 0.0159                                  & 0.1635 $\pm$ 0.0254                               & 1.9403 $\pm$ 0.5746                               \\
Chronos (Large)                      & 0.4280 $\pm$ 0.0649                                                    & 0.0960 $\pm$ 0.0244                                           & 0.0383 $\pm$ 0.0456                               & 1.0796 $\pm$ 0.2762                               \\
GPT4TS                               & \textbf{0.3022} $\pm$ 0.0206                                           & 0.1213 $\pm$ 0.0092                                           & 0.1534 $\pm$ 0.0618                               & 2.1992 $\pm$ 0.6330                               \\
DINN (ours)                          & 0.5656 $\pm$ \textbf{\color[HTML]{00009B} 0.0023}                               & 0.1391 $\pm$ \textbf{\color[HTML]{00009B} 0.0015}                      & \textbf{0.2772 $\pm$ \color[HTML]{00009B} 0.0035} & \textbf{4.4715 $\pm$ \color[HTML]{00009B} 0.0475} \\ \hline
\end{tabular}}
\captionsetup{font=footnotesize}
\caption{Comparative performance metrics for various time series models applied to the S\&P100 and DOW 30 dataset. Each entry represents the mean metric value along with the standard deviation. Metrics include Annualised Return (Ret), Annualised Standard Deviation (Std), Sharpe Ratio (SR), Sortino Ratio (SOR), Maximum Drawdown (MDD), Monthly 95\% Value-at-Risk (VaR), Return Over VaR (RoV), and accumulated terminal wealth (Wealth). Higher values are desirable for Ret, SR, SOR, RoV, and Wealth; while lower values are preferred for Std, MDD, and VaR. All values are presented as mean $\pm$ standard deviation across experimental trials. \textbf{Bold} values indicate the best performance for each metric, with upward ($\uparrow$) and downward ($\downarrow$) arrows indicating the desired direction of each measure. Values highlighted in \textcolor{blue}{blue} represent the lowest standard deviation for that metric.}
\label{tab1:main_result}
\vspace{-0.5cm}
\end{table}


\subsection{Why does prediction based loss function misalign with investment objectives?} 

A purely prediction-based loss function (e.g., minimizing mean-squared error) presumes that accurate forecasts of expected returns alone suffice for optimal investment decisions. In reality, portfolio optimization is highly sensitive to small forecast errors. Minor deviations in the predicted mean vector can lead to substantial misallocations of capital, especially when risk preferences and constraints amplify these inaccuracies.

Figure 1 illustrates this disconnect by comparing standard deviation outcomes for models trained only to minimize forecast error ("DFL w/o") versus those trained with an integrated decision module ("DFL w/"). Notably, the decision-informed approach exhibits a significantly lower standard deviation across experimental trials, signifying not only enhanced alignment with risk-return objectives but also greater robustness in performance. By aligning learned representations directly with portfolio-level goals, decision-informed approach may mitigates the volatility that often arises when small forecast errors are amplified within traditional MSE-based frameworks. Hence, reducing MSE does not always correlate with mitigating drawdowns, enhancing risk-adjusted returns, or boosting terminal wealth. The tighter variability achieved by the decision-focused model underscores that better forecasts do not necessarily translate into better investment outcomes. Instead, models must explicitly account for how forecast errors influence downstream allocation decisions in order to optimize both mean returns and risk exposure effectively.

\begin{figure}[h!] 
  \centering
  \includegraphics[width=\linewidth]{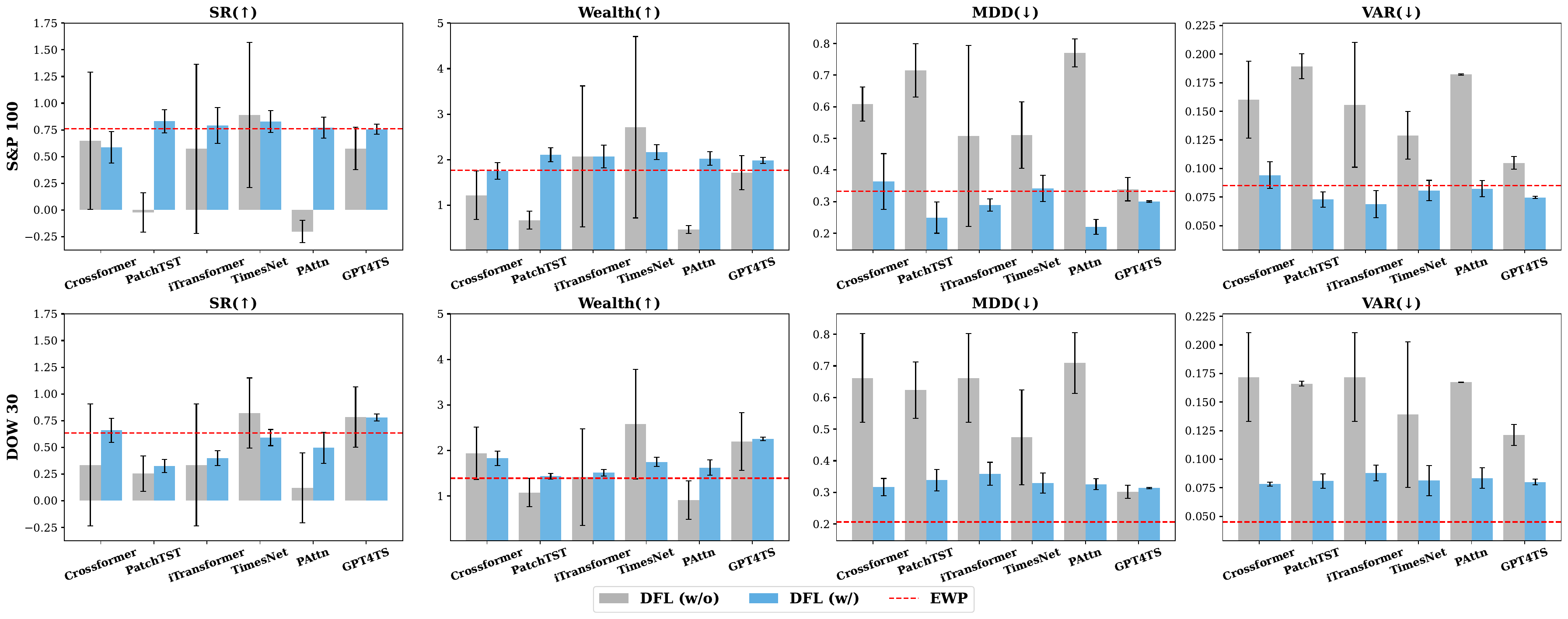}
   \captionsetup{font=footnotesize}
   \caption{Comparison of portfolio standard deviation across experimental trials for models trained with prediction-based loss only (DFL w/o) versus models incorporating a decision module (DFL w/). The decision-focused approach demonstrates notably lower variability in standard deviation outcomes, representing how integrating portfolio-level objectives during training leads to more consistent and robust investment performance compared to purely prediction-based optimization.}
  \label{fig:dfl}
\end{figure}

Proposition 1 provides a concrete theoretical example of this phenomenon. In a two-asset mean-variance problem ($\Sigma = I_2$ and $\lambda > 0$), we construct a sequence of predicted return vectors $\tilde{{\mu}}^{(k)}$ that converges to the true mean ${\mu}$ in $\ell_2$-norm (i.e., the MSE sense). Nonetheless, the induced optimal weights $\hat{{w}}$ do not converge to the true optimum ${w}^\star$. This result arises because mean-variance optimization can magnify small errors in the mean vector, thereby distorting the final portfolio solution. The implication is that standard predictive metrics—such as MSE—can overlook significant deviations in the resulting portfolio weights and performance.

\textbf{Proposition 1}: Let ${\mu}\in \mathbb{R}^n$ be the true expected return vector, and let $\hat{{\mu}} = \mathrm{arg\,min}_{{x} \in \mathcal{X}} \|{x} - {\mu}\|_{2}^{2}$.

Consider the mean-variance optimization problem
\begin{equation}
  {w}^{\star} = \mathrm{arg\,max}_{{w} \in \mathcal{W}}
  \{
    {w}^{\top}{\mu} - \lambda\,{w}^{\top}\Sigma\,{w}
  \},
\end{equation}
where $\lambda > 0$, $\Sigma \succ 0$, and $\mathcal{W} \subseteq \mathbb{R}^n$. Define similarly the portfolio
\begin{equation}
\hat{{w}} = \mathrm{arg\,max}_{{w} \in \mathcal{W}}
  \{
    {w}^{\top} \hat{{\mu}} -  \lambda\,{w}^{\top}\Sigma\,{w}
  \}.
\end{equation}
We show that there exist cases in which  
\begin{equation}
    \lim_{\|\hat{{\mu}} - {\mu}\|_2 \to 0} \hat{{w}}  \neq {w}^{\star}.
\end{equation}
In other words, even though $\hat{{\mu}}$ converges to ${\mu}$ in mean-squared error, the corresponding optimal portfolios need not converge to the true optimal portfolio ${w}^{\star}$. Consequently, a small MSE can lead to a large discrepancy in portfolio selection.

\textbf{Proof: } Consider the two-asset setting $(n = 2)$ with $\Sigma = I_2$ and $\lambda > 0$. Let the feasible set be
\begin{equation}
    \mathcal{W} = \{(w_1, w_2) : w_1 + w_2 = 1,  w_1, w_2 \ge 0 \}.
\end{equation}
Suppose the true return vector is ${\mu} = (\mu_1, \mu_2)^\top$ with $\mu_1 > \mu_2$.  Then the mean-variance optimization problem reduces to
\begin{equation}
    {w}^{\star} =\mathrm{arg\,max}_{\,w_1 + w_2 = 1,\;w_1,w_2 \ge 0}
\Bigl[
    w_1\,\mu_1+ w_2\,\mu_2 -\lambda\,\bigl(w_1^2 + w_2^2\bigr)
\Bigr].
\end{equation}

Since $w_2 = 1 - w_1$, define $   L(w_1) =  w_1\mu_1 + (1 - w_1)\mu_2 -\lambda \Bigl[w_1^2 + (1 - w_1)^2 \Bigr]$. Then, differentiating $L(w_1)$ and setting it to zero gives 
\begin{equation}
      \frac{dL}{dw_1} = \mu_1 -\mu_2 - \lambda\,\bigl[\,4\,w_1 - 2\bigr] = 0
\end{equation}

Solving for $w_1$ yields
\begin{equation}
    \mu_1 -\mu_2 =  \lambda\,[4w_1^\star - 2] \quad\Longrightarrow\quad w_1^\star =  \frac{1}{2} + \frac{\mu_1 - \mu_2}{4 \lambda}.
\end{equation}
Consequently, $w_2^\star = 1- w^{\star}_1$.

We then construct a specific sequence $\tilde{{\mu}}^{(k)}$ that converges to ${\mu}$ but whose induced portfolio weights fail to converge to ${w}^\star$. Set $\tilde{{\mu}}^{(k)} = \begin{pmatrix}  \mu_1 - \delta + \tfrac{1}{k}\\  \mu_2 \end{pmatrix}$ Where $\delta = \frac{\mu_1 - \mu_2}{2} > 0$. Clearly, as $k \to \infty$, $\tilde{{\mu}}^{(k)} \to {\mu}$ in $\ell_2$-norm because $\|\tilde{{\mu}}^{(k)} - {\mu}\|_2$ can be made arbitrarily small.

To show that the induced weights do not converge to ${w^{\star}}$, substitute $\tilde{{\mu}}^{(k)}$ into the same mean-variance formula. the optimal weight $w_{1}^{(k)}$ solves 
\begin{equation}
     (\mu_1 - \delta + \tfrac{1}{k}) - \mu_2 =    \lambda\ [ 4w_1^{(k)}- 2]
\end{equation}

Hence, $   w_1^{(k)} = \frac{1}{2} + \frac{(\mu_1 - \mu_2) - \delta +  \tfrac{1}{k}}{4\lambda}$.  By definition $\delta = \tfrac{\mu_1 - \mu_2}{2}$, we get
\begin{equation}
\begin{aligned}
w_1^{(k)} &= \frac{1}{2} + \frac{\tfrac{\mu_1 - \mu_2}{2} + \tfrac{1}{k}}{4\lambda},\\
          &= \frac{1}{2} + \frac{\mu_1 - \mu_2}{8\lambda} + \frac{1}{4\lambda k}.
\end{aligned}
\end{equation}
Meanwhile, the true optimum $w_1^\star$ is $w_1^\star = \frac{1}{2} + \frac{\mu_1 - \mu_2}{4\lambda}$. Observe that $\frac{\mu_1 - \mu_2}{8\lambda} + \frac{1}{4\lambda k} \neq \frac{\mu_1 - \mu_2}{4\lambda}$ So, $\lim_{k\to\infty} w_1^{(k)} \neq w_1^\star$. $\blacksquare$
\subsection{Can \texttt{DINN}'s attention mechanism enhance portfolio efficiency across varying market conditions?}

A central question in applying deep learning models to portfolio management is whether these models can systematically identify and emphasize assets that represent the market well while delivering favorable risk-adjusted returns under varying conditions. To explore this, we analyze the performance of \texttt{DINN} under four distinct macroeconomic regimes: the COVID-19 pandemic (March to June 2020), periods of elevated weekly initial jobless claims (ICSA), surges in new home sales (HSN1), and extremely low consumer sentiment (UMCS). During each regime, we evaluate four asset selection strategies: (1) \texttt{DINN}, using stocks deemed “important” by the prob-sparse attention mechanism, (2) Other, consisting of stocks not selected by the attention module, (3) Random, constructed with intentionally corrupted embeddings, and (4) Uniform, representing an equal-weighted portfolio.

\begin{table}[!htbp]
\centering
\centering
{\fontsize{8pt}{11pt}\selectfont
\begin{tabular}{cccccc}
\hline
\multicolumn{6}{l}{\textit{Panel A. S\&P 100 Dataset}}                                                              \\
Regime                 & Type        & SR($\uparrow$)  & Wealth($\uparrow$) & MDD($\downarrow$) & VAR($\downarrow$) \\ \hline
\multirow{4}{*}{COVID} & \texttt{DINN} & \textbf{1.1419} & \textbf{1.1576}    & \textbf{0.1915}   & 0.0342            \\
                       & Other       & 0.5857          & 1.0598             & 0.2829            & 0.0933            \\
                       & Random      & 0.6177          & 1.0096             & 0.2875            & \textbf{0.0330}   \\
                       & Uniform     & 0.6279          & 1.0683             & 0.2757            & 0.0883            \\ \hline
\multirow{4}{*}{ICSA}  & \texttt{DINN} & \textbf{1.5044} & \textbf{1.3299}    & \textbf{0.2408}   & \textbf{0.0527}   \\
                       & Other       & 0.8147          & 1.1502             & 0.2855            & 0.0822            \\
                       & Random      & 0.6896          & 1.0971             & 0.2609            & 0.0552            \\
                       & Uniform     & 0.9632          & 1.1887             & 0.2757            & 0.0755            \\ \hline
\multirow{4}{*}{HSN1}  & \texttt{DINN} & \textbf{0.6672} & \textbf{1.0672}    & 0.1688            & \textbf{0.0878}   \\
                       & Other       & 0.3937          & 1.0332             & \textbf{0.1651}   & 0.0941            \\
                       & Random      & 0.2394          & 0.9673             & 0.2996            & 0.1427            \\
                       & Uniform     & 0.4757          & 1.0429             & 0.1661            & 0.0923            \\ \hline
\multirow{4}{*}{UMCS}  & \texttt{DINN} & \textbf{2.2697} & \textbf{1.3505}    & 0.0908            & 0.0384            \\
                       & Other       & 2.0326          & 1.2937             & \textbf{0.0727}   & \textbf{0.0297}   \\
                       & Random      & 2.0096          & 1.1312             & 0.0890            & 0.0631            \\
                       & Uniform     & 2.1596          & 1.3166             & 0.0820            & 0.0332            \\ \hline
\multicolumn{6}{l}{}                                                                                                \\ \hline
\multicolumn{6}{l}{\textit{Panel B. DOW 30 Dataset}}                                                                \\
Regime                 & Type        & SR($\uparrow$)  & Wealth($\uparrow$) & MDD($\downarrow$) & VAR($\downarrow$) \\ \hline
\multirow{4}{*}{COVID} & \texttt{DINN} & \textbf{0.8742} & \textbf{1.0920}    & \textbf{0.2723}   & \textbf{0.0719}   \\
                       & Other       & 0.5830          & 1.0433             & 0.2845            & 0.0860            \\
                       & Random      & 0.6446          & 0.9651             & 0.3204            & 0.0724            \\
                       & Uniform     & 0.6511          & 1.0548             & 0.2798            & 0.0845            \\ \hline
\multirow{4}{*}{ICSA}  & \texttt{DINN} & 1.0319          & 1.2040             & 0.2669            & 0.0665            \\
                       & Other       & 0.9207          & 1.1676             & \textbf{0.2471}   & \textbf{0.0468}   \\
                       & Random      & 0.9822          & 1.1728             & 0.2915            & 0.0530            \\
                       & Uniform     & \textbf{1.0481} & \textbf{1.2108}    & 0.2798            & 0.0699            \\ \hline
\multirow{4}{*}{HSN1}  & \texttt{DINN} & \textbf{0.5812} & \textbf{1.0559}    & 0.1726            & \textbf{0.0902}   \\
                       & Other       & 0.3503          & 1.0303             & 0.2103            & 0.1037            \\
                       & Random      & 0.3565          & 1.0393             & 0.1839            & 0.0969            \\
                       & Uniform     & 0.5100          & 1.0461             & \textbf{0.1699}   & 0.0907            \\ \hline
\multirow{4}{*}{UMCS}  & \texttt{DINN} & \textbf{2.2088} & \textbf{1.3300}    & 0.0868            & 0.0338            \\
                       & Other       & 0.9519          & 1.1188             & 0.1004            & 0.0351            \\
                       & Random      & 1.6860          & 1.1814             & 0.0916            & \textbf{0.0326}   \\
                       & Uniform     & 2.1877          & 1.3206             & \textbf{0.0843}   & 0.0330            \\ \hline
\end{tabular}}

\captionsetup{font=footnotesize}
\caption{Performance comparison of portfolios constructed using different stock selection approaches across various market regimes from 2020 to 2023. \texttt{DINN} represents portfolios consisting of stocks selected by the prob-sparse attention mechanism, Other comprises stocks not selected by the attention mechanism, Random uses intentionally corrupted embedding information, and Uniform represents equal-weighted portfolios. Market regimes include: COVID-19 pandemic (March-June 2020), elevated initial jobless claims (ICSA $\ge$ 100,000, March-August 2020), housing market expansion (HSN1 $\ge$ 500, June-November 2022), and low consumer sentiment (UMCS $\le$ 60, May-December 2022). Performance metrics include Maximum Drawdown (MDD), Value at Risk (VaR), Sharpe Ratio (SR), and terminal wealth (Wealth). Arrows indicate whether lower ($\downarrow$) or higher ($\uparrow$) values are preferred. Bold values represent the best performance for each metric within each regime. Panel A reports results for the S\&P 100 dataset, while Panel B shows results for the DOW 30 dataset.}
\label{tab2:attns_result}
\vspace{-0.5cm}
\end{table}

\Cref{tab2:attns_result} presents the results across the S\&P 100 (Panel A) and DOW 30 (Panel B) datasets. The Sharpe ratios (SR) provide an intriguing perspective on \texttt{DINN}’s performance, suggesting that the attention mechanism may prioritize assets that efficiently balance risk and return. For instance, during the COVID-19 period, \texttt{DINN}-selected portfolios achieve an SR of 1.14 for the S\&P 100 and 0.87 for the DOW 30, exceeding the SR of portfolios formed from non-selected stocks (0.59 and 0.58, respectively). From an investment opportunity perspective\cite{kim2014cost}, these higher SR values could indicate that the attention mechanism identifies assets that approximate the efficient frontier more closely, potentially allowing for a more effective replication of market dynamics with fewer assets. In contrast, portfolios based on Random embeddings perform comparably to Uniform portfolios, with Sharpe ratios clustering close to those of simple equal-weighted strategies. This result appears to suggest that when embeddings are corrupted, the attention mechanism may lose its ability to prioritize meaningful assets effectively, leading to portfolios that do not achieve the same level of risk-adjusted returns observed with \texttt{DINN}.  

Other metrics, such as maximum drawdown (MDD) and terminal wealth, provide further observations. For example, in the ICSA regime, \texttt{DINN}-selected portfolios demonstrate lower MDD and higher terminal wealth compared to other strategies. These patterns suggest that \texttt{DINN}’s attention mechanism may adaptively select assets to mitigate downside risks while maintaining portfolio growth across diverse conditions. So, the results may indicate that \texttt{DINN}’s attention mechanism could contribute to constructing portfolios that more closely approximate the efficient frontier by selecting assets that represent the market effectively.

\subsection{How do we interpret $\frac{\partial \hat{w}_{t+h}}{\partial \hat{\mu}_{t+h}}$ and $\frac{\partial \hat{w}_{t+h}}{\partial \hat{\L}_{t+h}}$ within \texttt{DINN}'s portfolio decisions?}

Having established in Section 4.4 that \texttt{DINN}’s attention mechanism successfully isolates stocks of high importance, we now investigate whether such “importance” translates into an unintuitive performance gain in predictive accuracy. Specifically, we focus on two gradient-based sensitivities: (i) $\partial \hat{w}_{t+h}/\partial \hat{\mu}_{t+h}$ from Theorem 1, measuring how small changes in predicted returns $\hat{\mu}_{t+h}$ affect the model’s optimal weights $\hat{w}_{t+h}$, and (ii) $\partial \hat{w}/\partial \hat{L}$ from Theorem 2, measuring how the Cholesky factor $\hat{L}_{t+h}$ (and thus the predicted covariance $\hat{\Sigma}_{t+h}$) impacts $\hat{w}_{t+h}$. In principle, one might expect that assets whose weights are highly sensitive to errors in $\hat{\mu}_{t+h}$ or $\hat{\Sigma}_{t+h}$ (i.e., large gradients) would be more challenging to forecast, thereby yielding higher mean squared error (MSE). However, our findings contradict this conventional wisdom. When using only prediction loss, large gradients typically indicate a need for more model updates or suggest difficult-to-predict behavior. However, when incorporating decision-focused loss, we observe that these high-sensitivity assets actually show lower MSE. This occurs because DINN allocates greater learning capacity to stocks where prediction errors would result in higher decision-related costs. As a result, this improves predictive accuracy rather than increasing errors.

In \Cref{tab3:dwdmu_result}, we report the difference in MSE and MAE between “bottom” groups and “top” groups of assets, based on the absolute gradient magnitude ($\lvert\partial \hat{w}_{t+h}/\partial \hat{\mu}_{t+h}\rvert$). Specifically, the “10\%” column corresponds to $\bigr(\text{Bottom }10\% - \text{Top }10\%\bigr)$, meaning we first identify the 10\% of assets with the smallest gradients and the 10\% with the largest gradients, compute MSE or MAE for each group, and then subtract the top from the bottom. The same logic applies to the 20\% and 30\% columns. Panel A shows these differences for the S\&P 100 dataset, and Panel B for the DOW 30, spanning four macroeconomic regimes—COVID-19, ICSA, HSN1, UMCS—and the aggregated “ALL” period. For example, In Panel A, the S\&P 100 COVID-19 row have MSE of 2.4071 in the 10\% column means that the bottom-10\%-gradient group’s MSE is 2.4071 \emph{higher} than that of the top-10\%-gradient group. A similar pattern is evident across all regimes (ICSA, HSN1, UMCS) and is mirrored in MAE as well, consistently resulting in positive bottom-minus-top differences. Moving to the DOW 30 in Panel B, we observe the same phenomenon—for instance, the HSN1 regime shows a difference of 1.3214 in the 10\% column, implying the bottom group’s MSE exceeds that of the top group by 1.3214. 

We surmise that this arises because the decision-focused nature of \texttt{DINN} (and its training procedure) allocates additional modeling capacity to precisely those assets where misestimation would incur the highest decision-related costs. Consequently, \texttt{DINN} learns these “high-impact” assets more thoroughly, leading to lower MSE compared to stocks for which gradient-based sensitivities remain modest. Results for $\lvert \partial \hat{w}_{t+h}/\partial \hat{L}_{t+h} \rvert$ follow the same pattern (see Appendix A.4. for details).

\begin{table}[!htbp]
\centering
\centering
{\fontsize{7pt}{11pt}\selectfont
\begin{tabular}{ccccc}
\hline
\multicolumn{5}{l}{\textit{\textbf{Panel A. S\&P 100 Dataset}}}                                                                              \\
\multicolumn{1}{l}{\textbf{Regimes}} & \multicolumn{1}{l}{\textbf{Metric}} & \textbf{10\%}       & \textbf{20\%}       & \textbf{30\%}       \\ \hline
\multirow{2}{*}{COVID}               & MSE                                 & 2.4071 $\pm$ 0.0539 & 1.8473 $\pm$ 0.0064 & 1.3515 $\pm$ 0.0065 \\
                                     & MAE                                 & 0.4879 $\pm$ 0.0139 & 0.3680 $\pm$ 0.0004 & 0.2617 $\pm$ 0.0030 \\ \hline
\multirow{2}{*}{ICSA}                & MSE                                 & 2.5913 $\pm$ 0.0362 & 1.7386 $\pm$ 0.0042 & 1.2802 $\pm$ 0.0043 \\
                                     & MAE                                 & 0.5080 $\pm$ 0.0091 & 0.3546 $\pm$ 0.0003 & 0.2593 $\pm$ 0.0020 \\ \hline
\multirow{2}{*}{HSN1}                & MSE                                 & 0.7177 $\pm$ 0.0083 & 1.1894 $\pm$ 0.0064 & 1.0078 $\pm$ 0.0104 \\
                                     & MAE                                 & 0.2729 $\pm$ 0.0029 & 0.2877 $\pm$ 0.0025 & 0.2223 $\pm$ 0.0036 \\ \hline
\multirow{2}{*}{UMCS}                & MSE                                 & 2.4323 $\pm$ 0.0270 & 1.5520 $\pm$ 0.0034 & 1.2154 $\pm$ 0.0001 \\
                                     & MAE                                 & 0.5560 $\pm$ 0.0068 & 0.3573 $\pm$ 0.0021 & 0.2769 $\pm$ 0.0002 \\ \hline
\multirow{2}{*}{ALL}                 & MSE                                 & 1.7003 $\pm$ 0.0104 & 1.4622 $\pm$ 0.0045 & 1.1818 $\pm$ 0.0024 \\
                                     & MAE                                 & 0.4182 $\pm$ 0.0019 & 0.3401 $\pm$ 0.0008 & 0.2750 $\pm$ 0.0006 \\ \hline
\multicolumn{5}{l}{}                                                                                                                         \\ \hline
\multicolumn{5}{l}{\textit{\textbf{Panel B. DOW30 Dataset}}}                                                                                 \\
\textbf{Regimes}                     & \textbf{Metric}                     & \textbf{10\%}       & \textbf{20\%}       & \textbf{30\%}       \\ \hline
\multirow{2}{*}{COVID}               & MSE                                 & 0.6212 $\pm$ 0.0003 & 0.8646 $\pm$ 0.0005 & 0.8274 $\pm$ 0.0004 \\
                                     & MAE                                 & 0.1904 $\pm$ 0.0001 & 0.2545 $\pm$ 0.0002 & 0.2327 $\pm$ 0.0001 \\ \hline
\multirow{2}{*}{ICSA}                & MSE                                 & 0.6744 $\pm$ 0.0003 & 0.9512 $\pm$ 0.0002 & 0.8568 $\pm$ 0.0003 \\
                                     & MAE                                 & 0.2204 $\pm$ 0.0002 & 0.2846 $\pm$ 0.0001 & 0.2432 $\pm$ 0.0001 \\ \hline
\multirow{2}{*}{HSN1}                & MSE                                 & 1.3214 $\pm$ 0.0009 & 0.7704 $\pm$ 0.0007 & 0.7208 $\pm$ 0.0005 \\
                                     & MAE                                 & 0.4414 $\pm$ 0.0003 & 0.2478 $\pm$ 0.0002 & 0.2025 $\pm$ 0.0002 \\ \hline
\multirow{2}{*}{UMCS}                & MSE                                 & 0.9112 $\pm$ 0.0003 & 0.9274 $\pm$ 0.0005 & 0.8118 $\pm$ 0.0004 \\
                                     & MAE                                 & 0.3386 $\pm$ 0.0001 & 0.2970 $\pm$ 0.0001 & 0.2518 $\pm$ 0.0001 \\ \hline
\multirow{2}{*}{ALL}                 & MSE                                 & 1.1749 $\pm$ 0.0015 & 0.9416 $\pm$ 0.0042 & 0.8355 $\pm$ 0.0003 \\
                                     & MAE                                 & 0.3889 $\pm$ 0.0005 & 0.2884 $\pm$ 0.0013 & 0.2385 $\pm$ 0.0001 \\ \hline
\end{tabular}}
\captionsetup{font=footnotesize}
\caption{Differences in prediction error (\textit{Bottom}~\(x\%\)~\(-\)~\textit{Top}~\(x\%\)) for assets grouped by the absolute gradient $\bigl|\partial \hat{w}/\partial \hat{\mu}\bigr|$ under four macroeconomic regimes (COVID-19, ICSA, HSN1, UMCS) plus an aggregated “ALL” period. Panel A presents results for the S\&P 100 and Panel B for the DOW 30. The columns labeled 10\%, 20\%, and 30\% indicate the difference in mean squared error (MSE) or mean absolute error (MAE) between the bottom-$x\%$ group (smallest gradients) and the top-$x\%$ group (largest gradients). Each cell shows the average $\pm$ standard deviation computed across multiple training runs (random seeds). Positive values imply that high-gradient assets yield lower errors, suggesting that \texttt{DINN} prioritizes forecasting accuracy where decision-related costs are greatest.}
\label{tab3:dwdmu_result}
\vspace{-0.5cm}
\end{table}


\label{ch:ch5}
\section{Conclusion} This paper addresses a longstanding challenge in quantitative finance: bridging the gap between more accurate forecasts of financial variables and truly optimal portfolio decisions. While improved prediction accuracy is often cited as the path to superior investment returns, our empirical and theoretical findings demonstrate that purely predictive approaches can fail to yield the best portfolio outcomes. Drawing on recent developments in decision-focused learning, we proposed the \texttt{\texttt{DINN}} (Decision-Informed Neural Network) framework, which not only advances the state of the art in financial forecasting by incorporating large language models (LLMs) but also directly integrates a portfolio optimization layer into the end-to-end training process.

From an empirical standpoint, the experiments conducted on two representative equity datasets (S\&P 100 and DOW 30) suggest three key findings. First, \texttt{DINN} delivers systematically stronger performance across a broad set of metrics—including annualized return, Sharpe ratio, and terminal wealth—when compared to standard deep learning baselines, such as Transformer variants and other LLM-based architectures that rely solely on traditional prediction losses. Second, the inclusion of a prob-sparse attention mechanism may helps the model identify and emphasize a smaller subset of assets critical to replicating market dynamics under a variety of macroeconomic conditions. This mechanism not only focuses the model on economically significant information but also yields portfolios with lower drawdowns and higher risk-adjusted performance during stress regimes (e.g., the COVID-19 crisis and spikes in jobless claims). Third, the gradient-based sensitivity analyses provide a theoretical framework through which to interpret \texttt{DINN}’s asset allocations: \textit{high-sensitivity assets}, which would inflict larger “regret” if incorrectly predicted, exhibit lower mean-squared errors than less-sensitive assets. This finding underscores that \texttt{DINN} “learns what matters” by adjusting its representational power to minimize precisely those errors most detrimental to the ultimate portfolio objective.

Methodologically, the paper makes several contributions that advance the intersection of machine learning and portfolio optimization. It develops a rigorous pipeline to incorporate LLM representations of both inter-asset relationships (e.g., sector-level textual prompts) and macroeconomic data (e.g., summary embeddings of irregularly sampled indicators), thereby enriching the model’s feature space without overwhelming it with noise. By differentiating directly through a convex optimization layer, \texttt{DINN} closes the prediction-decision gap: improving return forecasts is no longer an end in itself but a means to more robust investment decisions.

Looking ahead, three avenues of future research emerge. First, while the current formulation centers on a mean-variance objective with convex risk constraints, extending decision-focused learning to alternative objectives—such as value-at-risk or expected shortfall—may further enhance real-world robustness. Second, although LLM-driven embeddings capture textual and structured macroeconomic signals, ongoing advances in multimodal data ingestion (e.g., satellite imagery or social media feeds) could further refine how markets’ evolving information sets are integrated into portfolio weights. Lastly, large-scale empirical analyses across broader asset classes, such as fixed income or commodities, would help validate and generalize the \texttt{DINN} approach beyond equity-centric portfolios.
In conclusion, this paper shows that bridging the divide between forecasting and portfolio choice requires going beyond optimizing for statistical accuracy alone. By merging representation learning and end-to-end differentiable optimization, \texttt{DINN} offers a systematic way to ensure that improvements in predictive modeling directly translate into meaningful gains in investment decisions. We hope this framework will serve as a foundation for future work in decision-focused learning for finance, spurring both theoretical advances in differentiable optimization techniques and innovative empirical applications across various market settings.



\bibliographystyle{ref_style}
\bibliography{ref_data}

\begin{thebibliography}{63}
\providecommand{\natexlab}[1]{#1}
\providecommand{\noopsort}[1]{}
\providecommand{\printfirst}[2]{#1}
\providecommand{\singleletter}[1]{#1}
\providecommand{\switchargs}[2]{#2#1}

\bibitem[\protect\citeauthoryear{Achiam {\itshape{et~al.}}}{2023}]{achiam2023gpt}
Achiam, J., Adler, S., Agarwal, S., Ahmad, L., Akkaya, I., Aleman, F.L., Almeida, D., Altenschmidt, J., Altman, S., Anadkat, S. {\itshape et~al.}, Gpt-4 technical report. {\itshape arXiv preprint arXiv:2303.08774}, 2023.

\bibitem[\protect\citeauthoryear{Agrawal {\itshape{et~al.}}}{2019}]{cvxpylayers2019}
Agrawal, A., Amos, B., Barratt, S., Boyd, S., Diamond, S. and Kolter, Z., Differentiable Convex Optimization Layers. In {\itshape Proceedings of the }{\itshape Advances in Neural Information Processing Systems}, 2019.

\bibitem[\protect\citeauthoryear{Amos and Kolter}{2017}]{amos2017optnet}
Amos, B. and Kolter, J.Z., Optnet: Differentiable optimization as a layer in neural networks. In {\itshape Proceedings of the }{\itshape International conference on machine learning}, pp. 136--145, 2017.

\bibitem[\protect\citeauthoryear{Anis and Kwon}{2025}]{anis2025end}
Anis, H.T. and Kwon, R.H., End-to-end, decision-based, cardinality-constrained portfolio optimization. {\itshape European Journal of Operational Research}, 2025, \textbf{320}, 739--753.

\bibitem[\protect\citeauthoryear{Ansari {\itshape{et~al.}}}{2024}]{ansari2024chronos}
Ansari, A.F., Stella, L., Turkmen, C., Zhang, X., Mercado, P., Shen, H., Shchur, O., Rangapuram, S.S., Arango, S.P., Kapoor, S. {\itshape et~al.}, Chronos: Learning the Language of Time Series. {\itshape Transactions on Machine Learning Research}, 2024.

\bibitem[\protect\citeauthoryear{Ban {\itshape{et~al.}}}{2018}]{ban2018machine}
Ban, G.Y., El~Karoui, N. and Lim, A.E., Machine learning and portfolio optimization. {\itshape Management Science}, 2018, \textbf{64}, 1136--1154.

\bibitem[\protect\citeauthoryear{Bekaert {\itshape{et~al.}}}{2002}]{bekaert2002dynamics}
Bekaert, G., Harvey, C.R. and Lumsdaine, R.L., The dynamics of emerging market equity flows. {\itshape Journal of International money and Finance}, 2002, \textbf{21}, 295--350.

\bibitem[\protect\citeauthoryear{Best and Grauer}{1991}]{best1991sensitivity}
Best, M.J. and Grauer, R.R., On the sensitivity of mean-variance-efficient portfolios to changes in asset means: some analytical and computational results. {\itshape The review of financial studies}, 1991, \textbf{4}, 315--342.

\bibitem[\protect\citeauthoryear{Butler and Kwon}{2023}]{butler2023integrating}
Butler, A. and Kwon, R.H., Integrating prediction in mean-variance portfolio optimization. {\itshape Quantitative Finance}, 2023, \textbf{23}, 429--452.

\bibitem[\protect\citeauthoryear{Cao {\itshape{et~al.}}}{2024}]{cao2024tempo}
Cao, D., Jia, F., Arik, S.O., Pfister, T., Zheng, Y., Ye, W. and Liu, Y., {TEMPO}: Prompt-based Generative Pre-trained Transformer for Time Series Forecasting. In {\itshape Proceedings of the }{\itshape The Twelfth International Conference on Learning Representations}, 2024.

\bibitem[\protect\citeauthoryear{Cenesizoglu and Timmermann}{2012}]{cenesizoglu2012return}
Cenesizoglu, T. and Timmermann, A., Do return prediction models add economic value?. {\itshape Journal of Banking \& Finance}, 2012, \textbf{36}, 2974--2987.

\bibitem[\protect\citeauthoryear{Chan {\itshape{et~al.}}}{1999}]{chan1999portfolio}
Chan, L.K., Karceski, J. and Lakonishok, J., On portfolio optimization: Forecasting covariances and choosing the risk model. {\itshape The review of Financial studies}, 1999, \textbf{12}, 937--974.

\bibitem[\protect\citeauthoryear{Chen {\itshape{et~al.}}}{2024}]{chen2024deep}
Chen, L., Pelger, M. and Zhu, J., Deep learning in asset pricing. {\itshape Management Science}, 2024, \textbf{70}, 714--750.

\bibitem[\protect\citeauthoryear{Chopra and Ziemba}{1993}]{chopra1993effect}
Chopra, V.K. and Ziemba, W.T., The effect of errors in means, variances, and covariances on optimal portfolio choice. {\itshape Journal of Portfolio Management}, 1993, \textbf{19}, 6.

\bibitem[\protect\citeauthoryear{Chung {\itshape{et~al.}}}{2022}]{chung2022effects}
Chung, M., Lee, Y., Kim, J.H., Kim, W.C. and Fabozzi, F.J., The effects of errors in means, variances, and correlations on the mean-variance framework. {\itshape Quantitative Finance}, 2022, \textbf{22}, 1893--1903.

\bibitem[\protect\citeauthoryear{Costa and Iyengar}{2023}]{costa2023distributionally}
Costa, G. and Iyengar, G.N., Distributionally robust end-to-end portfolio construction. {\itshape Quantitative Finance}, 2023, \textbf{23}, 1465--1482.

\bibitem[\protect\citeauthoryear{DeMiguel {\itshape{et~al.}}}{2009}]{demiguel2009optimal}
DeMiguel, V., Garlappi, L. and Uppal, R., Optimal versus naive diversification: How inefficient is the 1/N portfolio strategy?. {\itshape The review of Financial studies}, 2009, \textbf{22}, 1915--1953.

\bibitem[\protect\citeauthoryear{Dubey {\itshape{et~al.}}}{2024}]{dubey2024llama}
Dubey, A., Jauhri, A., Pandey, A., Kadian, A., Al-Dahle, A., Letman, A., Mathur, A., Schelten, A., Yang, A., Fan, A. {\itshape et~al.}, The llama 3 herd of models. {\itshape arXiv preprint arXiv:2407.21783}, 2024.

\bibitem[\protect\citeauthoryear{Elmachtoub and Grigas}{2022}]{elmachtoub2022smart}
Elmachtoub, A.N. and Grigas, P., Smart “predict, then optimize”. {\itshape Management Science}, 2022, \textbf{68}, 9--26.

\bibitem[\protect\citeauthoryear{Fama and French}{1993}]{fama1993common}
Fama, E.F. and French, K.R., Common risk factors in the returns on stocks and bonds. {\itshape Journal of financial economics}, 1993, \textbf{33}, 3--56.

\bibitem[\protect\citeauthoryear{Fama and French}{2015}]{fama2015five}
Fama, E.F. and French, K.R., A five-factor asset pricing model. {\itshape Journal of financial economics}, 2015, \textbf{116}, 1--22.

\bibitem[\protect\citeauthoryear{Feng {\itshape{et~al.}}}{2020}]{feng2020taming}
Feng, G., Giglio, S. and Xiu, D., Taming the factor zoo: A test of new factors. {\itshape The Journal of Finance}, 2020, \textbf{75}, 1327--1370.

\bibitem[\protect\citeauthoryear{Firoozye {\itshape{et~al.}}}{2023}]{firoozye2023canonical}
Firoozye, N., Tan, V. and Zohren, S., Canonical portfolios: Optimal asset and signal combination. {\itshape Journal of Banking \& Finance}, 2023, \textbf{154}, 106952.

\bibitem[\protect\citeauthoryear{Gerber {\itshape{et~al.}}}{2022}]{gerber2022gerber}
Gerber, S., Markowitz, H.M., Ernst, P.A., Miao, Y., Javid, B. and Sargen, P., The Gerber Statistic: A Robust Co-Movement Measure for Portfolio Optimization.. {\itshape Journal of Portfolio Management}, 2022, \textbf{48}.

\bibitem[\protect\citeauthoryear{Giglio {\itshape{et~al.}}}{2022}]{giglio2022factor}
Giglio, S., Kelly, B. and Xiu, D., Factor models, machine learning, and asset pricing. {\itshape Annual Review of Financial Economics}, 2022, \textbf{14}, 337--368.

\bibitem[\protect\citeauthoryear{Gu {\itshape{et~al.}}}{2020}]{gu2020empirical}
Gu, S., Kelly, B. and Xiu, D., Empirical asset pricing via machine learning. {\itshape The Review of Financial Studies}, 2020, \textbf{33}, 2223--2273.

\bibitem[\protect\citeauthoryear{Guidolin and Timmermann}{2007}]{guidolin2007asset}
Guidolin, M. and Timmermann, A., Asset allocation under multivariate regime switching. {\itshape Journal of Economic Dynamics and Control}, 2007, \textbf{31}, 3503--3544.

\bibitem[\protect\citeauthoryear{Hwang {\itshape{et~al.}}}{2024}]{hwang2024temporal}
Hwang, Y., Zohren, S. and Lee, Y., Temporal Representation Learning for Stock Similarities and Its Applications in Investment Management. {\itshape arXiv preprint arXiv:2407.13751}, 2024.

\bibitem[\protect\citeauthoryear{Jagannathan and Ma}{2003}]{jagannathan2003risk}
Jagannathan, R. and Ma, T., Risk reduction in large portfolios: Why imposing the wrong constraints helps. {\itshape The journal of finance}, 2003, \textbf{58}, 1651--1683.

\bibitem[\protect\citeauthoryear{Jin {\itshape{et~al.}}}{2024}]{jin2023time}
Jin, M., Wang, S., Ma, L., Chu, Z., Zhang, J.Y., Shi, X., Chen, P.Y., Liang, Y., Li, Y.F., Pan, S. and Wen, Q., {Time-LLM}: Time series forecasting by reprogramming large language models. In {\itshape Proceedings of the }{\itshape International Conference on Learning Representations (ICLR)}, 2024.

\bibitem[\protect\citeauthoryear{Kelly {\itshape{et~al.}}}{2019}]{kelly2019characteristics}
Kelly, B.T., Pruitt, S. and Su, Y., Characteristics are covariances: A unified model of risk and return. {\itshape Journal of Financial Economics}, 2019, \textbf{134}, 501--524.

\bibitem[\protect\citeauthoryear{Kim {\itshape{et~al.}}}{2021{\natexlab{a}}}]{kim2021mean}
Kim, J.H., Lee, Y., Kim, W.C. and Fabozzi, F.J., Mean-variance optimization for asset allocation. {\itshape Journal of Portfolio Management}, 2021{\natexlab{a}}, \textbf{47}, 24--40.

\bibitem[\protect\citeauthoryear{Kim {\itshape{et~al.}}}{2024}]{kim2024overview}
Kim, J.H., Lee, Y., Kim, W.C., Kang, T. and Fabozzi, F.J., An Overview of Optimization Models for Portfolio Management. {\itshape Journal of Portfolio Management}, 2024, \textbf{51}.

\bibitem[\protect\citeauthoryear{Kim {\itshape{et~al.}}}{2021{\natexlab{b}}}]{kim2021reversible}
Kim, T., Kim, J., Tae, Y., Park, C., Choi, J.H. and Choo, J., Reversible instance normalization for accurate time-series forecasting against distribution shift. In {\itshape Proceedings of the }{\itshape International Conference on Learning Representations}, 2021{\natexlab{b}}.

\bibitem[\protect\citeauthoryear{Kim {\itshape{et~al.}}}{2014}]{kim2014cost}
Kim, W.C., Lee, Y. and Lee, Y.H., Cost of Asset Allocation in Equity Market: How Much Do Investors Lose Due to Bad Asset Class Design?. {\itshape The Journal of Portfolio Management}, 2014, \textbf{41}, 34--44.

\bibitem[\protect\citeauthoryear{Kourtis {\itshape{et~al.}}}{2012}]{kourtis2012parameter}
Kourtis, A., Dotsis, G. and Markellos, R.N., Parameter uncertainty in portfolio selection: Shrinking the inverse covariance matrix. {\itshape Journal of Banking \& Finance}, 2012, \textbf{36}, 2522--2531.

\bibitem[\protect\citeauthoryear{Ledoit and Wolf}{2003}]{ledoit2003improved}
Ledoit, O. and Wolf, M., Improved estimation of the covariance matrix of stock returns with an application to portfolio selection. {\itshape Journal of empirical finance}, 2003, \textbf{10}, 603--621.

\bibitem[\protect\citeauthoryear{Ledoit and Wolf}{2004}]{ledoit2004well}
Ledoit, O. and Wolf, M., A well-conditioned estimator for large-dimensional covariance matrices. {\itshape Journal of multivariate analysis}, 2004, \textbf{88}, 365--411.

\bibitem[\protect\citeauthoryear{Lee {\itshape{et~al.}}}{2024}]{lee2024overview}
Lee, Y., Kim, J.H., Kim, W.C. and Fabozzi, F.J., An Overview of Machine Learning for Portfolio Optimization.. {\itshape Journal of Portfolio Management}, 2024, \textbf{51}.

\bibitem[\protect\citeauthoryear{Lintner}{1975}]{lintner1975valuation}
Lintner, J., The valuation of risk assets and the selection of risky investments in stock portfolios and capital budgets. In {\itshape Stochastic optimization models in finance}, pp. 131--155, 1975, Elsevier.

\bibitem[\protect\citeauthoryear{Liu {\itshape{et~al.}}}{2024}]{liu2023itransformer}
Liu, Y., Hu, T., Zhang, H., Wu, H., Wang, S., Ma, L. and Long, M., iTransformer: Inverted Transformers Are Effective for Time Series Forecasting. {\itshape The Twelfth International Conference on Learning Representations}, 2024.

\bibitem[\protect\citeauthoryear{L{\"o}ffler}{2003}]{loffler2003effects}
L{\"o}ffler, G., The effects of estimation error on measures of portfolio credit risk. {\itshape Journal of Banking \& Finance}, 2003, \textbf{27}, 1427--1453.

\bibitem[\protect\citeauthoryear{Mandi {\itshape{et~al.}}}{2024}]{mandi2024decision}
Mandi, J., Kotary, J., Berden, S., Mulamba, M., Bucarey, V., Guns, T. and Fioretto, F., Decision-focused learning: Foundations, state of the art, benchmark and future opportunities. {\itshape Journal of Artificial Intelligence Research}, 2024.

\bibitem[\protect\citeauthoryear{Markowitz}{1952}]{Markowitz1952}
Markowitz, H., Portfolio Selection. {\itshape The Journal of Finance}, 1952, \textbf{7}, 77--91.

\bibitem[\protect\citeauthoryear{Michaud}{1989}]{michaud1989markowitz}
Michaud, R.O., The Markowitz optimization enigma: Is ‘optimized’optimal?. {\itshape Financial analysts journal}, 1989, \textbf{45}, 31--42.

\bibitem[\protect\citeauthoryear{Nie {\itshape{et~al.}}}{2023}]{Yuqietal_PatchTST}
Nie, Y., H.~Nguyen, N., Sinthong, P. and Kalagnanam, J., A Time Series is Worth 64 Words: Long-term Forecasting with Transformers. In {\itshape Proceedings of the }{\itshape International Conference on Learning Representations}, 2023.

\bibitem[\protect\citeauthoryear{Nie {\itshape{et~al.}}}{2024}]{nie2024survey}
Nie, Y., Kong, Y., Dong, X., Mulvey, J.M., Poor, H.V., Wen, Q. and Zohren, S., A Survey of Large Language Models for Financial Applications: Progress, Prospects and Challenges. {\itshape arXiv preprint arXiv:2406.11903}, 2024.

\bibitem[\protect\citeauthoryear{Petersen {\itshape{et~al.}}}{2008}]{petersen2008matrix}
Petersen, K.B., Pedersen, M.S. {\itshape et~al.}, The matrix cookbook. {\itshape Technical University of Denmark}, 2008, \textbf{7}, 510.

\bibitem[\protect\citeauthoryear{Romanko {\itshape{et~al.}}}{2023}]{romanko2023chatgpt}
Romanko, O., Narayan, A. and Kwon, R.H., Chatgpt-based investment portfolio selection. In {\itshape Proceedings of the }{\itshape Operations Research Forum}, Vol. ~4, p.~91, 2023.

\bibitem[\protect\citeauthoryear{Rousseeuw and Driessen}{1999}]{rousseeuw1999fast}
Rousseeuw, P.J. and Driessen, K.V., A fast algorithm for the minimum covariance determinant estimator. {\itshape Technometrics}, 1999, \textbf{41}, 212--223.

\bibitem[\protect\citeauthoryear{Sharpe}{1964}]{sharpe1964capital}
Sharpe, W.F., Capital asset prices: A theory of market equilibrium under conditions of risk. {\itshape The journal of finance}, 1964, \textbf{19}, 425--442.

\bibitem[\protect\citeauthoryear{Tan {\itshape{et~al.}}}{2024}]{tan2024language}
Tan, M., Merrill, M.A., Gupta, V., Althoff, T. and Hartvigsen, T., Are language models actually useful for time series forecasting?. In {\itshape Proceedings of the }{\itshape The Thirty-eighth Annual Conference on Neural Information Processing Systems}, 2024.

\bibitem[\protect\citeauthoryear{Tan and Zohren}{2020}]{tan2020estimation}
Tan, V. and Zohren, S., Estimation of Large Financial Covariances: A Cross-Validation Approach. {\itshape arXiv preprint arXiv:2012.05757}, 2020.

\bibitem[\protect\citeauthoryear{Van~Aelst and Rousseeuw}{2009}]{van2009minimum}
Van~Aelst, S. and Rousseeuw, P., Minimum volume ellipsoid. {\itshape Wiley Interdisciplinary Reviews: Computational Statistics}, 2009, \textbf{1}, 71--82.

\bibitem[\protect\citeauthoryear{Waswani {\itshape{et~al.}}}{2017}]{waswani2017attention}
Waswani, A., Shazeer, N., Parmar, N., Uszkoreit, J., Jones, L., Gomez, A., Kaiser, L. and Polosukhin, I., Attention is all you need. In {\itshape Proceedings of the }{\itshape NIPS}, 2017.

\bibitem[\protect\citeauthoryear{Wu {\itshape{et~al.}}}{2023}]{wu2023timesnet}
Wu, H., Hu, T., Liu, Y., Zhou, H., Wang, J. and Long, M., TimesNet: Temporal 2D-Variation Modeling for General Time Series Analysis. In {\itshape Proceedings of the }{\itshape International Conference on Learning Representations}, 2023.

\bibitem[\protect\citeauthoryear{Wu {\itshape{et~al.}}}{2021}]{wu2021autoformer}
Wu, H., Xu, J., Wang, J. and Long, M., Autoformer: Decomposition Transformers with {Auto-Correlation} for Long-Term Series Forecasting. In {\itshape Proceedings of the }{\itshape Advances in Neural Information Processing Systems}, 2021.

\bibitem[\protect\citeauthoryear{Zhang {\itshape{et~al.}}}{2021}]{zhang2021universal}
Zhang, C., Zhang, Z., Cucuringu, M. and Zohren, S., A universal end-to-end approach to portfolio optimization via deep learning. {\itshape arXiv preprint arXiv:2111.09170}, 2021.

\bibitem[\protect\citeauthoryear{Zhang and Yan}{2023}]{zhang2023crossformer}
Zhang, Y. and Yan, J., Crossformer: Transformer Utilizing Cross-Dimension Dependency for Multivariate Time Series Forecasting. In {\itshape Proceedings of the }{\itshape International Conference on Learning Representations}, 2023.

\bibitem[\protect\citeauthoryear{Zhou {\itshape{et~al.}}}{2021}]{informer_2021}
Zhou, H., Zhang, S., Peng, J., Zhang, S., Li, J., Xiong, H. and Zhang, W., Informer: Beyond Efficient Transformer for Long Sequence Time-Series Forecasting. In {\itshape Proceedings of the }{\itshape The Thirty-Fifth {AAAI} Conference on Artificial Intelligence, {AAAI} 2021, Virtual Conference}, Vol. ~35, pp. 11106--11115, 2021, {AAAI} Press.

\bibitem[\protect\citeauthoryear{Zhou {\itshape{et~al.}}}{2022}]{zhou2022fedformer}
Zhou, T., Ma, Z., Wen, Q., Wang, X., Sun, L. and Jin, R., {FEDformer}: Frequency enhanced decomposed transformer for long-term series forecasting. In {\itshape Proceedings of the }{\itshape Proc. 39th International Conference on Machine Learning (ICML 2022)}, Baltimore, Maryland, 2022.

\bibitem[\protect\citeauthoryear{Zhou {\itshape{et~al.}}}{2023{\natexlab{a}}}]{zhou2023one}
Zhou, T., Niu, P., Sun, L., Jin, R. {\itshape et~al.}, One fits all: Power general time series analysis by pretrained lm. {\itshape Advances in neural information processing systems}, 2023{\natexlab{a}}, \textbf{36}, 43322--43355.

\bibitem[\protect\citeauthoryear{Zhou {\itshape{et~al.}}}{2023{\natexlab{b}}}]{zhou2023onefitsall}
Zhou, T., Niu, P., Sun, L., Jin, R. {\itshape et~al.}, {One Fits All}: Power General Time Series Analysis by Pretrained LM. In {\itshape Proceedings of the }{\itshape NeurIPS}, 2023{\natexlab{b}}.

\end{thebibliography}

\clearpage
\label{ch:appendix}
\section*{Appendix}
\section*{Appendix A.1.: Detailed Proofs and Derivations} \label{Appendix:A1}

\noindent\textbf{Proof of Theorem 1.} \label{Appendix:A1}

\textbf{Proof: } Consider the problem \Cref{eq:modified_problem}. Introducing Lagrange multipliers $\eta$ and $\gamma$ for the risk and budget constraints, respectively, the Lagrangian is here: 
\begin{equation}
    \mathcal{L}(w,s_{t+h},\eta,\gamma)=\lambda s_{t+h}^2 - \hat{\mu}_{t+h}^{\top} \hat{w}_{t+h} + \eta(\hat{w}_{t+h}^{\top}\hat{\Sigma}_{t+h} \hat{w}_{t+h} - s_{t+h}) + \gamma(\mathbf{1}^{\top} \hat{w}_{t+h} -1)
\end{equation}
Differentiating with respect to $s_{t+h}$ gives $2\lambda s_{t+h}-\eta =0 \implies \eta=2\lambda s_{t+h}.$ Differentiation with respect to $w$ then yields $-\hat{\mu}_{t+h} + 2\lambda \hat{\Sigma}_{t+h} \hat{w}_{t+h} + \gamma \mathbf{1}=0.$ Hence $\hat{\mu}_{t+h} = 2\lambda \hat{\Sigma}_{t+h} \hat{w}_{t+h} + \gamma \mathbf{1}$. Since $\mathbf{1}^{\top}w=1,$ we have $w = \tfrac{1}{2\lambda}\hat{\Sigma}_{t+h}^{-1}(\hat{\mu}_{t+h}-\gamma\mathbf{1}).$ Multiplying by $\mathbf{1}^{\top}$ and solving for $\gamma$ gives $\gamma = \tfrac{\mathbf{1}^{\top}\hat{\Sigma}_{t+h}^{-1}\hat{\mu}_{t+h}-2\lambda}{\mathbf{1}^{\top}\hat{\Sigma}_{t+h}^{-1}\mathbf{1}}.$  Substituting back into the expression for $w$ and setting $2\lambda=1$ (which does not affect the structure of the solution) yields
\begin{equation}
    w = \hat{\Sigma}_{t+h}^{-1}\hat{\mu}_{t+h}-\hat{\Sigma}_{t+h}^{-1}\frac{\mathbf{1}^{\top}\hat{\Sigma}_{t+h}^{-1}\hat{\mu}_{t+h}-1}{\mathbf{1}^{\top}\hat{\Sigma}_{t+h}^{-1}\mathbf{1}}\mathbf{1}
\end{equation}
Differentiating this with respect to $\hat{\mu}_{t+h}$ and simplifying leads to
\begin{equation}
    \frac{\partial \hat{w}_{t+h}}{\partial \hat{\mu}_{t+h}} = \hat{\Sigma}_{t+h}^{-1}-\frac{\hat{\Sigma}_{t+h}^{-1}\mathbf{1}\mathbf{1}^{\top}\hat{\Sigma}_{t+h}^{-1}}{\mathbf{1}^{\top}\hat{\Sigma}_{t+h}^{-1}\mathbf{1}},
\end{equation}
as claimed $\blacksquare$.

\noindent\textbf{Proof of Theorem 2.} \label{Appendix:A2}

\textbf{Proof: } Starting from the expression derived in Theorem 1 under the normalization $2\lambda=1$, the optimal portfolio weights can be written as  $\hat{w}_{t+h} = \hat{\Sigma}_{t+h}^{-1}\hat{\mu}_{t+h} - p\,\hat{\Sigma}_{t+h}^{-1}\mathbf{1}$, where  $p = \frac{\mathbf{1}^{\top}\hat{\Sigma}_{t+h}^{-1}\hat{\mu}_{t+h} - 1}{\mathbf{1}^{\top}\hat{\Sigma}_{t+h}^{-1}\mathbf{1}}$.  
In this formulation, $\hat{\Sigma}_{t+h}^{-1}$ depends on $\hat{L}_{t+h}$ through the relation $\hat{\Sigma}_{t+h} = \hat{L}_{t+h}\hat{L}_{t+h}^{\top}$. Thus, the chain rule of differentiation implies that understanding $\partial \hat{w}_{t+h}/\partial \hat{\Sigma}_{t+h}$ allows recovery of $\partial \hat{w}_{t+h}/\partial \hat{L}_{t+h}$.

Differentiating $\hat{w}_{t+h}$ with respect to $\hat{\Sigma}_{t+h}$ first requires considering the terms $\hat{\Sigma}_{t+h}^{-1}\hat{\mu}_{t+h}$ and $\hat{\Sigma}_{t+h}^{-1}\mathbf{1}$. From standard matrix calculus \citep{petersen2008matrix}, one has $\partial \hat{\Sigma}^{-1}/\partial \hat{\Sigma} = -\hat{\Sigma}^{-1}(\cdot)\hat{\Sigma}^{-1}$. Applying this to $\hat{\Sigma}_{t+h}^{-1}\hat{\mu}_{t+h}$ yields $\frac{\partial (\hat{\Sigma}_{t+h}^{-1}\hat{\mu}_{t+h})}{\partial \hat{\Sigma}_{t+h}} = -\hat{\Sigma}_{t+h}^{-1}\hat{\mu}_{t+h}\hat{\Sigma}_{t+h}^{-1}$.

Similarly, the term involving $p$ is more involved since $p$ itself depends on $\hat{\Sigma}_{t+h}^{-1}$. Letting $g=\mathbf{1}^{\top}\hat{\Sigma}_{t+h}^{-1}\hat{\mu}_{t+h}$ and $z=\mathbf{1}^{\top}\hat{\Sigma}_{t+h}^{-1}\mathbf{1}$, one has $p=(g-1)/z$. Differentiating $g$ and $z$ with respect to $\hat{\Sigma}_{t+h}$ gives  
\begin{equation}
    \frac{\partial g}{\partial \hat{\Sigma}_{t+h}} = -\hat{\Sigma}_{t+h}^{-1}\mathbf{1}\hat{\mu}_{t+h}^{\top}\hat{\Sigma}_{t+h}^{-1}, \quad
\frac{\partial z}{\partial \hat{\Sigma}_{t+h}} = -\hat{\Sigma}_{t+h}^{-1}\mathbf{1}\mathbf{1}^{\top}\hat{\Sigma}_{t+h}^{-1}.
\end{equation}

Applying the quotient rule to differentiate $p=(g-1)/z$ yields  
\begin{equation}
    \frac{\partial p}{\partial \hat{\Sigma}_{t+h}}
= \frac{-\hat{\Sigma}_{t+h}^{-1}\mathbf{1}\hat{\mu}_{t+h}^{\top}\hat{\Sigma}_{t+h}^{-1}z + (g-1)\hat{\Sigma}_{t+h}^{-1}\mathbf{1}\mathbf{1}^{\top}\hat{\Sigma}_{t+h}^{-1}}{z^2}
\end{equation}

Combining these results, the derivative of $p\,\hat{\Sigma}_{t+h}^{-1}\mathbf{1}$ with respect to $\hat{\Sigma}_{t+h}$ is  
\begin{equation}
    \frac{\partial (p\,\hat{\Sigma}_{t+h}^{-1}\mathbf{1})}{\partial \hat{\Sigma}_{t+h}}
= \frac{\partial p}{\partial \hat{\Sigma}_{t+h}}\hat{\Sigma}_{t+h}^{-1}\mathbf{1} - p\,\hat{\Sigma}_{t+h}^{-1}\mathbf{1}\hat{\Sigma}_{t+h}^{-1}
\end{equation}

Subtracting this from $-\hat{\Sigma}_{t+h}^{-1}\hat{\mu}_{t+h}\hat{\Sigma}_{t+h}^{-1}$ and rearranging terms leads to  
\begin{equation}
    \frac{\partial \hat{w}_{t+h}}{\partial \hat{\Sigma}_{t+h}}
= -\hat{\Sigma}_{t+h}^{-1}(\hat{\mu}_{t+h}-p\mathbf{1})\hat{\Sigma}_{t+h}^{-1} - \frac{\partial p}{\partial \hat{\Sigma}_{t+h}}\hat{\Sigma}_{t+h}^{-1}\mathbf{1}
\end{equation}

Since $\hat{\Sigma}_{t+h} = \hat{L}_{t+h}\hat{L}_{t+h}^{\top}$, differentiating with respect to $\hat{L}_{t+h}$ involves applying the chain rule. Under appropriate vectorization, symmetry assumptions, the lower-triangular structure of $\hat{L}_{t+h}$, and considering only independent parameters, the derivative $\partial \hat{\Sigma}_{t+h}/\partial \hat{L}_{t+h}$ can be simplified to contribute a factor of $2\hat{L}_{t+h}$. Substituting back, the final result is  
\begin{equation}
\begin{aligned}
\frac{\partial \hat{w}_{t+h}}{\partial \hat{L}_{t+h}} 
&= \frac{\partial \hat{w}_{t+h}}{\partial \hat{\Sigma}_{t+h}}\frac{\partial \hat{\Sigma}_{t+h}}{\partial \hat{L}_{t+h}}, \\
&= -2\,\hat{\Sigma}_{t+h}^{-1}(\hat{\mu}_{t+h}-p\mathbf{1})\hat{\Sigma}_{t+h}^{-1}\hat{L}_{t+h} 
- 2\,\hat{\Sigma}_{t+h}^{-1}\left(\frac{\partial p}{\partial \hat{\Sigma}_{t+h}}\mathbf{1}\right)\hat{L}_{t+h}
\end{aligned}
\end{equation} $\blacksquare$.

\section*{Appendix A.3: Hyper-Parameter Configuration Details} \label{Appendix:A3}

All experiments in this paper were conducted on a workstation with four NVIDIA RTX 3090 GPUs. Unless otherwise noted, the training proceeded with a mini-batch size of 16. Below, we detail the key hyper-parameter ranges and selection criteria employed for model training and evaluation. The codes are available at \href{https://anonymous.4open.science/r/Decision-informed-Neural-Networks-with-Large-Language-Model-Integration-for-Portfolio-Optimization-A441/README.md}{Anonymous Github}.

\subsection*{Model Hyper-Parameters and training strategy's}
\begin{itemize}
\item \textbf{Attention Heads:} We examined configurations with either 2 or 4 attention heads. Preliminary experiments indicated that increasing attention heads can help capture more nuanced inter-asset relationships; however, larger numbers of heads also slightly increase computational cost.
\item \textbf{Encoder Depths:} We explored encoder depths of 1, 2, and 4 layers. Deeper encoders generally improved representational capacity, albeit at the risk of overfitting if not adequately regularized.
\item \textbf{LLM Hidden Dimensions:} To reduce model size while preserving performance, we tested hidden dimensions of 12, 24, 36, and 72 for the Large Language Model (LLM) backbone. These smaller dimensions (compared to standard large LLM deployments) were sufficient for the financial time-series tasks in this paper and allowed us to balance model expressiveness with training efficiency.
\item \textbf{Training Epochs and Early Stopping:} All models were trained up to a maximum of 50 epochs. We employed early stopping on a validation set to prevent overfitting, monitoring the composite loss (prediction + decision-focused components) for convergence.
\item \textbf{Optimizer and Learning Rate:} We used the Adam optimizer with a base learning rate of $1 \times 10^{-4}$. Additionally, we adopted the following dynamic learning rate adjustment strategy \citep{jin2023time}.

\end{itemize}

\subsection*{Portfolio Optimization Settings}
For the decision-focused optimization component, we selected the risk-aversion parameter $\lambda$ from a candidate set $\{0.0145,\ 0.2656,\ 0.9545,\ 2.4305,\ 3.4623\}$. These five values respectively correspond to portfolios that may be characterized as highly aggressive, aggressive, balanced, conservative and highly conservative.
The final $\lambda$ used throughout the main text was $0.9545$, which corresponds to the “balanced” risk profile. For completeness, \Cref{ap1:abl_risk} compares the out-of-sample portfolio performance under these various risk-aversion levels.

\setcounter{table}{0}
\renewcommand{\thetable}{A.\arabic{table}}
\begin{table}[!htbp]
\centering
\centering
{\fontsize{8pt}{11pt}\selectfont
\begin{tabular}{lcccc}
\hline
\multicolumn{5}{l}{\textit{\textbf{Panel A. S\&P 100 Dataset}}}                                                                                                  \\
\multicolumn{1}{c}{\textbf{Measure}} & \textbf{Ret ($\uparrow$)}    & \textbf{Std ($\downarrow$)}  & \textbf{SR ($\uparrow$)}     & \textbf{SOR ($\uparrow$)}    \\ \hline
Highly Aggressive                    & 0.4099 $\pm$ 0.0031          & 0.4220 $\pm$ 0.0025          & 0.9446 $\pm$ 0.0103          & 1.3699 $\pm$ 0.0175          \\
Aggressive                           & 0.4128 $\pm$ 0.0299          & 0.4113 $\pm$ 0.0024          & 0.9762 $\pm$ 0.0686          & 1.4130 $\pm$ 0.1054          \\
Balanced                             & \textbf{0.4353 $\pm$ 0.0145} & 0.4103 $\pm$ 0.0005          & \textbf{1.0335 $\pm$ 0.0358} & \textbf{1.5008 $\pm$ 0.0521} \\
Conservative                         & 0.4024 $\pm$ 0.0050          & 0.4021 $\pm$ 0.0013          & 0.9727 $\pm$ 0.0098          & 1.3912 $\pm$ 0.0170          \\
Highly Conservative                  & 0.3266 $\pm$ 0.0069          & \textbf{0.3912 $\pm$ 0.0014} & 0.8062 $\pm$ 0.0182          & 1.1450 $\pm$ 0.0306          \\ \hline
\multicolumn{1}{c}{\textbf{Measure}} & \textbf{MDD ($\downarrow$)}  & \textbf{VaR ($\downarrow$)}  & \textbf{RoV ($\uparrow$)}    & \textbf{Welath ($\uparrow$)} \\ \hline
Highly Aggressive                    & 0.4048 $\pm$ 0.0000          & 0.1230 $\pm$ 0.0000          & 0.1843 $\pm$ 0.0019          & 2.7580 $\pm$ 0.0294          \\
Aggressive                           & 0.3989 $\pm$ 0.0012          & 0.1230 $\pm$ 0.0000          & 0.1875 $\pm$ 0.0146          & 2.8355 $\pm$ 0.2289          \\
Balanced                             & 0.3951 $\pm$ 0.0064          & 0.1233 $\pm$ 0.0004          & \textbf{0.1987 $\pm$ 0.0078} & \textbf{3.0213 $\pm$ 0.1218} \\
Conservative                         & 0.3849 $\pm$ 0.0039          & 0.1252 $\pm$ 0.0012          & 0.1822 $\pm$ 0.0036          & 2.7891 $\pm$ 0.0349          \\
Highly Conservative                  & \textbf{0.3841 $\pm$ 0.0128} & \textbf{0.1212 $\pm$ 0.0011} & 0.1520 $\pm$ 0.0050          & 2.2733 $\pm$ 0.0479          \\ \hline
\multicolumn{5}{l}{}                                                                                                                                             \\ \hline
\multicolumn{5}{l}{\textit{\textbf{Panel B. DOW 30 Dataset}}}                                                                                                    \\
\multicolumn{1}{c}{\textbf{Measure}} & \textbf{Ret ($\uparrow$)}    & \textbf{Std ($\downarrow$)}  & \textbf{SR ($\uparrow$)}     & \textbf{SOR ($\uparrow$)}    \\ \hline
Highly Aggressive                    & \textbf{0.6512 $\pm$ 0.0143} & 0.4891 $\pm$ 0.0004          & 1.3085 $\pm$ 0.0283          & 1.9925 $\pm$ 0.0427          \\
Aggressive                           & 0.6389 $\pm$ 0.0070          & 0.4860 $\pm$ 0.0003          & 1.2915 $\pm$ 0.0146          & 1.9527 $\pm$ 0.0215          \\
Balanced                             & 0.6325 $\pm$ 0.0043          & 0.4814 $\pm$ 0.0002          & 1.2905 $\pm$ 0.0091          & 1.9449 $\pm$ 0.0137          \\
Conservative                         & 0.6394 $\pm$ 0.0143          & 0.4722 $\pm$ 0.0014          & \textbf{1.3302 $\pm$ 0.0337} & \textbf{2.0170 $\pm$ 0.0475} \\
Highly Conservative                  & 0.4987 $\pm$ 0.0058          & \textbf{0.4466 $\pm$ 0.0023} & 1.0914 $\pm$ 0.0125          & 1.6039 $\pm$ 0.0198          \\ \hline
\multicolumn{1}{c}{\textbf{Measure}} & \textbf{MDD ($\downarrow$)}  & \textbf{VaR ($\downarrow$)}  & \textbf{RoV ($\uparrow$)}    & \textbf{Welath ($\uparrow$)} \\ \hline
Highly Aggressive                    & 0.5745 $\pm$ 0.0013          & 0.1591 $\pm$ 0.0000          & 0.2498 $\pm$ 0.0051          & 3.6121 $\pm$ 0.1572          \\
Aggressive                           & 0.5759 $\pm$ 0.0012          & 0.1570 $\pm$ 0.0036          & 0.2476 $\pm$ 0.0061          & 3.5024 $\pm$ 0.0773          \\
Balanced                             & 0.5656 $\pm$ 0.0023          & 0.1391 $\pm$ 0.0015          & 0.2772 $\pm$ 0.0035          & \textbf{3.4715 $\pm$ 0.0475} \\
Conservative                         & \textbf{0.5446 $\pm$ 0.0101} & 0.1240 $\pm$ 0.0036          & \textbf{0.3148 $\pm$ 0.0150} & 3.6288 $\pm$ 0.1700          \\
Highly Conservative                  & 0.5567 $\pm$ 0.0056          & \textbf{0.1224 $\pm$ 0.0006} & 0.2525 $\pm$ 0.0025          & 3.3819 $\pm$ 0.0500          \\ \hline
\end{tabular}}

\captionsetup{font=footnotesize}
\caption{Comparative performance metrics for various time series models applied to the S\&P100 and DOW 30 dataset. Each entry represents the mean metric value along with the standard deviation. Metrics include Annualised Return (Ret), Annualised Standard Deviation (Std), Sharpe Ratio (SR), Sortino Ratio (SOR), Maximum Drawdown (MDD), Monthly 95\% Value-at-Risk (VaR), Return Over VaR (RoV), and accumulated terminal wealth (Wealth). Higher values are desirable for Ret, SR, SOR, RoV, and Wealth; while lower values are preferred for Std, MDD, and VaR. All values are presented as mean $\pm$ standard deviation across experimental trials. \textbf{Bold} values indicate the best performance for each metric, with upward ($\uparrow$) and downward ($\downarrow$) arrows indicating the desired direction of each measure.}
\label{ap1:abl_risk}
\vspace{-0.5cm}
\end{table}

\clearpage
\section*{Appendix A.4: Additional Gradient-Based Analysis for $\boldsymbol{\hat{L}_{t+h}}$} \label{Appendix:A5}

Appendix K provides a more extensive examination of how $\bigl|\partial \hat{w}_{t+h} / \partial \hat{L}_{t+h}\bigr|$ influences forecast accuracy under a decision-focused neural network framework. In the main text, Section 4.5 focuses on gradient-based sensitivities for the predicted mean, $\hat{\mu}_{t+h}$. Here, we address corresponding sensitivities for $\hat{L}_{t+h}$, the Cholesky factor of the predicted covariance $\hat{\Sigma}_{t+h}$. Conceptually, assets whose allocations are highly sensitive to $\hat{L}_{t+h}$ (that is, those exhibiting large magnitudes for $\partial \hat{w}_{t+h} / \partial \hat{L}_{t+h}$) should receive more modeling “effort,” since inaccurate estimation of their covariance structure could prove costly for downstream portfolio decisions. If a model is truly decision-focused, it will tend to reduce errors specifically for high-sensitivity assets, thereby securing improved overall portfolio performance.

To explore this phenomenon, we replicate the same grouping strategy used in Section 4.5. Results in \Cref{tab:abs_dwdl} consistently exhibit positive values for both MSE and MAE differences across most market regimes and for both datasets. This indicates that assets with higher gradient magnitudes, which the portfolio optimization layer deems more influential for risk management, experience smaller forecasting errors than do lower-sensitivity assets. For instance, focusing on the S\&P 100 COVID row in Panel A, the difference of 2.3696 for MSE in the 10\% column means that the bottom group’s mean squared error is 2.3696 points larger than that of the top group. The same type of result characterizes other regimes, such as ICSA, HSN1, and UMCS, as well as the ALL category that aggregates the entire test period. A parallel pattern arises in the DOW 30 data, reinforcing the same conclusion: the bottom group (in terms of $\bigl|\partial \hat{w}_{t+h} / \partial \hat{L}_{t+h}\bigr|$) is forecast less accurately than the top group. Observing larger differences supports the notion that the decision-focused approach dedicates more learnable parameters or training “focus” to stocks whose covariance-factor misestimation would most detrimentally affect the risk-return trade-off.

\setcounter{table}{2}
\renewcommand{\thetable}{A.\arabic{table}}
\begin{table}[!htbp]
\centering
\centering
{\fontsize{8pt}{11pt}\selectfont
\begin{tabular}{ccccc}
\hline
\multicolumn{5}{l}{\textit{\textbf{Panel A. S\&P 100 Dataset}}}                                                                              \\
\textbf{Regimes}                     & \textbf{Metric}                     & \textbf{10\%}       & \textbf{20\%}       & \textbf{30\%}       \\ \hline
COVID                                & MSE                                 & 2.3696 $\pm$ 0.0009 & 1.8748 $\pm$ 0.0029 & 1.3597 $\pm$ 0.0002 \\
COVID                                & MAE                                 & 0.4779 $\pm$ 0.0002 & 0.3783 $\pm$ 0.0001 & 0.2659 $\pm$ 0.0002 \\ \hline
ICSA                                 & MSE                                 & 2.5663 $\pm$ 0.0009 & 1.7569 $\pm$ 0.0018 & 1.2723 $\pm$ 0.0002 \\
ICSA                                 & MAE                                 & 0.5014 $\pm$ 0.0003 & 0.3615 $\pm$ 0.0001 & 0.2589 $\pm$ 0.0001 \\ \hline
HSN1                                 & MSE                                 & 0.7049 $\pm$ 0.0008 & 1.1256 $\pm$ 0.0782 & 1.0007 $\pm$ 0.0064 \\
HSN1                                 & MAE                                 & 0.2688 $\pm$ 0.0002 & 0.2757 $\pm$ 0.0120 & 0.2197 $\pm$ 0.0019 \\ \hline
UMCS                                 & MSE                                 & 2.4424 $\pm$ 0.0008 & 1.5530 $\pm$ 0.0043 & 1.2054 $\pm$ 0.0001 \\
UMCS                                 & MAE                                 & 0.5583 $\pm$ 0.0003 & 0.3589 $\pm$ 0.0020 & 0.2745 $\pm$ 0.0002 \\ \hline
ALL                                  & MSE                                 & 1.6990 $\pm$ 0.0144 & 1.4595 $\pm$ 0.0119 & 1.1908 $\pm$ 0.0012 \\
ALL                                  & MAE                                 & 0.4169 $\pm$ 0.0022 & 0.3414 $\pm$ 0.0026 & 0.2758 $\pm$ 0.0001 \\ \hline
\multicolumn{5}{l}{}                                                                                                                         \\ \hline
\multicolumn{5}{l}{\textit{\textbf{Panel B. DOW30 Dataset}}}                                                                                 \\
\multicolumn{1}{c}{\textbf{Regimes}} & \multicolumn{1}{c}{\textbf{Metric}} & \textbf{10\%}       & \textbf{20\%}       & \textbf{30\%}       \\ \hline
\multicolumn{1}{c}{COVID}            & \multicolumn{1}{c}{MSE}             & 0.6212 $\pm$ 0.0003 & 0.8727 $\pm$ 0.0005 & 0.8256 $\pm$ 0.0003 \\
\multicolumn{1}{c}{COVID}            & \multicolumn{1}{c}{MAE}             & 0.1904 $\pm$ 0.0001 & 0.2614 $\pm$ 0.0002 & 0.2326 $\pm$ 0.0002 \\ \hline
\multicolumn{1}{c}{ICSA}             & \multicolumn{1}{c}{MSE}             & 0.6891 $\pm$ 0.0003 & 0.9493 $\pm$ 0.0004 & 0.8556 $\pm$ 0.0003 \\
\multicolumn{1}{c}{ICSA}             & \multicolumn{1}{c}{MAE}             & 0.2271 $\pm$ 0.0002 & 0.2864 $\pm$ 0.0002 & 0.2430 $\pm$ 0.0001 \\ \hline
\multicolumn{1}{c}{HSN1}             & \multicolumn{1}{c}{MSE}             & 1.1017 $\pm$ 0.0008 & 0.7566 $\pm$ 0.0007 & 0.7534 $\pm$ 0.0299 \\
\multicolumn{1}{c}{HSN1}             & \multicolumn{1}{c}{MAE}             & 0.3740 $\pm$ 0.0003 & 0.2406 $\pm$ 0.0002 & 0.1962 $\pm$ 0.0038 \\ \hline
\multicolumn{1}{c}{UMCS}             & \multicolumn{1}{c}{MSE}             & 0.9322 $\pm$ 0.0003 & 0.9362 $\pm$ 0.0004 & 0.8109 $\pm$ 0.0004 \\
\multicolumn{1}{c}{UMCS}             & \multicolumn{1}{c}{MAE}             & 0.3446 $\pm$ 0.0001 & 0.3007 $\pm$ 0.0001 & 0.2517 $\pm$ 0.0001 \\ \hline
\multicolumn{1}{c}{ALL}              & \multicolumn{1}{c}{MSE}             & 1.1638 $\pm$ 0.0052 & 0.9219 $\pm$ 0.0006 & 0.8351 $\pm$ 0.0032 \\
\multicolumn{1}{c}{ALL}              & \multicolumn{1}{c}{MAE}             & 0.3839 $\pm$ 0.0005 & 0.2809 $\pm$ 0.0003 & 0.2383 $\pm$ 0.0004 \\ \hline
\end{tabular}}
\captionsetup{font=footnotesize}
\caption{Differences in prediction error (\textit{Bottom}~$x\%$~$-$~\textit{Top}~$x\%$) for assets grouped by the absolute gradient $\bigl\lvert \partial \hat{w}_{t+h}/\partial \hat{\mu}_{t+h}\bigr\rvert$ under four macroeconomic regimes (COVID-19, ICSA, HSN1, UMCS) plus an aggregated “ALL” period. Panel A presents results for the S\&P 100 and Panel B for the DOW 30. The columns labeled 10\%, 20\%, and 30\% indicate the difference in mean squared error (MSE) or mean absolute error (MAE) between the bottom-$x\%$ group (smallest gradients) and the top-$x\%$ group (largest gradients). Each cell shows the average $\pm$ standard deviation computed across multiple training runs (random seeds). Positive values imply that high-gradient assets yield lower errors, suggesting that \texttt{DINN} prioritizes forecasting accuracy where decision-related costs are greatest.}
\label{tab:abs_dwdl}
\vspace{-0.5cm}
\end{table}


\end{document}